\documentclass[useAMS,usenatbib]{mn2e}
\usepackage[usenames,dvipsnames]{color}
\usepackage{amssymb,mn2e-breakabs,graphicx}
\usepackage{natbib}
\usepackage{subfigure}

\begin{document}

\title[The evolution of quiescent galaxies at high redshifts ( z $\geq$ 1.4)]{The evolution of quiescent galaxies at high redshifts ( z $\geq$ 1.4)} 
\author[H. Dom{\'i}nguez S{\'a}nchez et al.]{H. Dom{\'i}nguez S{\'a}nchez$^{1,2}$ \thanks{E-mail:
helena.dominguez@oabo.inaf.it}, F.~Pozzi$^3$, C.~Gruppioni$^1$, A.~Cimatti$^3$, O.~Ilbert$^4$,
 \newauthor
L.~Pozzetti$^1$, H.~McCracken$^5$, P.~Capak$^6$, E.~ Le Floch$^7$, M.~Salvato$^8$, G.~Zamorani$^1$,
\newauthor 
 C.~M.~Carollo$^9$,
 T.~Contini$^{10}$,
 J.-P.~Kneib$^{11}$,
 O.~Le~F\`evre$^{11}$,
 S.~J.~Lilly$^{9}$,
 V.~Mainieri$^{12}$,
\newauthor
 A.~Renzini$^{13}$,
 M.~Scodeggio$^{14}$,
 S.~Bardelli$^1$,
 M.~Bolzonella$^1$,
 A.~Bongiorno$^{15}$,
 K.~Caputi$^{16,9}$,
\newauthor
 G.~Coppa$^{1,3}$,
 O.~Cucciati$^{17}$,
 S.~de~la~Torre$^{14,18}$,
 L.~de~Ravel$^{11,16}$,
 P.~Franzetti$^{14}$,
 B.~Garilli$^{14}$,
\newauthor 
A.~Iovino$^{18}$,
 P.~Kampczyk$^9$,
 C.~Knobel$^9$,
 K.~Kova\v{c}$^9$,
 F.~Lamareille$^{10}$,
 J.-F.~Le~Borgne$^{10}$,
\newauthor
 V.~Le Brun$^{11}$,
 C.~Maier$^9$,
 M.~Mignoli$^1$,
 R.~Pell\'o$^{10}$,
 Y.~Peng$^9$,
 E.~Perez-Montero$^{10,19}$,
\newauthor
 E.~Ricciardelli$^{13,2}$,
 J.~D.~Silverman$^{9,20}$,
 M.~Tanaka$^{12,20}$,
 L.~A.~M.~Tasca$^{11,14}$,
 L.~Tresse$^{11}$,
\newauthor
 D.~Vergani$^{1}$, 
 E.~Zucca$^{1}$
\\
$^{1}$INAF-Osservatorio Astronomico di Bologna, via Ranzani 1, I-40127 Bologna, Italy\\
$^{2}$Instituto de Astrof{\'i}sica de Canarias, 38205 La Laguna, Spain\\
$^{3}$Alma Mater Studiorum Universit\`a di Bologna, Dipartimento di Astronomia, via Ranzani 1, I-40127 Bologna, Italy\\
$^{4}$Laboratoire d'astrophysic de Marseille, Universit\`{e} de provence, CNRS,BP 8,Traverse du Siphon,13376 Marseille Cedex 12, France\\
$^{5}$Institute d'astrophysic de Paris, UMR7095 CNRS, Universit\`{e} Pierre et Marie Curie, 98 bis Boulevard Arago, 75014 Paris, France\\
$^{6}$Spitzer Science Center, 314-6 Caltech, Pasadena, CA 91125.; 105-24 Caltech, Pasadena, CA 91125\\
$^{7}$Laboratoire AIM, CEA/DSM-CNRS-Universite Paris Diderot, IRFU/Service d'Astrophysique, CEA-Saclay, 91191 Gif-sur-Yvette Cedex, France\\
$^{8}$Max Planck Institute for Plasma Physics and Excellence Cluster Universe, Boltzmannstr. 2 D-85748 Garching, Germany\\
$^{9}$ETH Zurich, Institute of Astronomy, Wolfgang-Pauli-Stra{\ss}e 27, 8093 Zurich, Switzerland\\
$^{10}$Laboratoire d'Astrophysique de Toulouse-Tarbes, Universit\'{e} de Toulouse, CNRS, 14 avenue Edouard Belin, 31400 Toulouse, France\\ 
$^{11}$Laboratoire d'Astrophysique de Marseille, Universit\'{e} d'Aix-Marseille, CNRS, 38 rue Fr\'{e}d\'{e}ric Joliot-Curie, 13388 Marseille Cedex 13, France\\ 
$^{12}$European Southern Observatory, Karl-Schwarzschild-Stra{\ss}e 2, 85748 Garching bei M\"unchen, Germany\\ 
$^{13}$INAF - Osservatorio Astronomico di Padova, vicolo dell'Osservatorio 5, 35122 Padova, Italy\\ 
$^{14}$INAF - IASF Milano, via Bassini 15, 20133 Milano, Italy \\
$^{15}$Max-Planck-Institut f\"ur Extraterrestrische Physik, Giessenbachstra{\ss}e, 84571 Garching bei M\"unchen, Germany\\
$^{16}$Institute for Astronomy, Royal Observatory, Blackford Hill, Edinburgh, EH9 3HJ, Scotland, United Kingdom\\ 
$^{17}$INAF - Osservatorio Astronomico di Trieste, via G. B. Tiepolo 11, 34143 Trieste, Italy\\ 
$^{18}$INAF - Osservatorio Astronomico di Brera, via Brera 28, 20121 Milano, Italy\\ 
$^{19}$Instituto de Astrofisica de Andalucia, CSIC, Apdo. 3004, 18080 Granada, Spain\\
$^{20}$Institute for the Physics and Mathematics of the Universe (IPMU), University of Tokyo, Kashiwanoha 5-1-5, Kashiwa-shi, Chiba 277-8568, Japan\\ 
}

\date{
Accepted 2011 June 15. Received 2011 June 15; in original form 2011 March 30}

\maketitle 
 

\begin{abstract}

The goal of this work is to study the evolution of high redshift ($z \geq
1.4$) quiescent galaxies over an effective area of $\sim 1.7$ deg$^{2}$ in the COSMOS
field. Galaxies have been divided according to their star-formation activity and
the evolution of the different populations, in particular of the quiescent galaxies, has been investigated in detail.
 We have studied an IRAC ($ mag_{3.6 \mu m}< $22.0) selected sample of $\sim$ 18000 
galaxies at z $\geq$ 1.4 in the COSMOS field with multi-wavelength coverage
extending from the \textit{U} band to the \textit{Spitzer} 24 $\mu$m one.  We have derived
accurate photometric redshifts ($\sigma_{\Delta z/(1+z_{s})}=0.06$) through a
SED-fitting procedure. Other important physical parameters (masses, ages and
star formation rates (SFR)) of the galaxies have been obtained using Maraston
(2005) models. We have divided our sample into actively star-forming,
intermediate and quiescent galaxies depending on their specific star formation
rate ($SSFR=SFR/M$). We have computed the galaxy stellar mass function (GSMF) of
the total  sample and the different populations at $z=1.4-3.0$. 
We have studied the properties of high redshift quiescent galaxies finding that they are old ($1-4$ Gyr), massive ($\langle M \rangle \sim
10^{10.65}$M$_{\odot}$), weakly star forming stellar populations with low dust
extinction (E(B-V) $\leq$ 0.15) and small e-folding time scales ($\tau \sim0.1-0.3$ Gyr).
 We observe a significant evolution of the quiescent stellar mass function from $2.5 < z < 3.0$ to $1.4 < z < 1.6$,
 increasing by $\sim 1$ dex in this redshift interval. 
We find that  $z \sim 1.5$ is an epoch of transition of the GSMF: while the GSMF at
 $z \gtrsim 1.5$ is dominated by the star-forming galaxies at all stellar masses,
 at $z \lesssim 1.5$ the contribution to the total GSMF of the quiescent galaxies is
 significant and becomes higher than that of the star-forming population
for $M \geqslant 10^{10.75}$M$_{\odot}$. 
The fraction of star-forming galaxies decreases from $60\%$ to $20\%$ from $z\sim 2.5-3.0$  to
$z\sim $ $1.4-1.6$ for $M \sim 10^{11.0}$M$_{\odot}$, while the quiescent population
increases from $10\%$ to $50\%$ at the same redshift and mass intervals. We
compare the fraction of quiescent galaxies derived with that predicted by
theoretical models and find that the Kitzbichler $\&$ White (2007) model, implemented on the Millennium Simulation,
 is the one that better reproduces the shape of the data.
 Finally, we calculate the stellar mass density of the star-forming and quiescent populations as a function of redshift and find that
there is already a significant number of quiescent galaxies at $z > 2.5$ 
($\rho \sim 6.0$ M$_{\odot}$Mpc$^{-3}$), meaning that efficient star formation had
to take place before that time.
 

\end{abstract}

\section{Introduction}

Understanding the processes that regulate stellar mass growth in galaxies as
well as tracing the history of galaxy star formation and mass assembly over the
cosmic time (\citealt{Lilly1996}, \citealt{Madau1996}, \citealt{Pozzetti1998},
\citealt{Dickinson2003}) are among the most important and discussed topics in
modern cosmology.
During the last decade, extraordinary progress has been made in our
comprehension of the formation and evolution of galaxies. Thanks to deep
multi-wavelength surveys we are now able to observe large samples of distant
galaxies, thus reducing cosmic variance and allowing us to study objects at
early cosmic times.

In the present Universe, it is known that galaxies can be separated into two broad populations: 
passively evolving galaxies with red colors, also known as quiescent galaxies or early type 
galaxies (ETGs hereafter), and star-forming galaxies with blue colors (e.g., \citealt{Baldry2004},
\citealt{Brinchmann2004}). A similar division extends to at least $ z \sim 1$ (\citealt{Bell2004},
\citealt{Willmer2006}) and possibly to $z \sim 2$ (\citealt{Giallongo2005}, \citealt{Cirasuolo2007}, \citealt{Cassata2008}).

ETGs are the most massive galaxies in the present day Universe and are characterized by
simple and homogeneous properties (morphologies, colors, passively evolving stellar populations,
 scaling relations). As a consequence, they are a key population to investigate the stellar mass assembly 
of massive galaxies over the cosmic time (e.g., \citealt{Renzini2006}).
\citet{Gallazzi2006} studied the properties of local quiescent galaxies and, in particular, the color-magnitude
and the Mg$_{2}$-$\sigma_{v}$ relation. They found that for high-mass elliptical galaxies the dispersion in age is small, 
while at the low mass end there is a tail towards younger ages, reflecting a shift in stellar growth towards
less massive galaxies in recent epochs. Besides, they also argue that at increasing stellar mass there is also an increase 
in both total metallicity and $\alpha/$Fe ratio, suggesting that massive early type galaxies formed in a relatively short time-scale.
The stellar mass and ages of the passive ETGs at $z\sim 1-2$ require precursors
characterized by strong ($ > 100$M$_{\odot}/$yr) and short-lived ($0.1-0.3$ Gyr) starbursts occurring at $z > 2-3$ (e.g., \citealt{Cimatti2009} 
and references therein). 

Stellar mass assembly in galaxies subdivided by spectral and morphological
types, as well as by star formation activity has been investigated at $z> 0$ by many authors
(e.g., \citealt{Bundy2005}, \citealt{Franceschini2006}, \citealt{Panella2006}, \citealt{Pozzetti2010}).
\citet{Cimatti2006} studied the B-band luminosity function of ETGs since $z \sim 1$, finding that the amount of the evolution for the ETG
 population depends critically on the range of masses considered, i.e., most of the massive ETGs are already in place at $z\thicksim$1, while the
density is still increasing with time for lower masses. This means that more massive galaxies
have older stellar populations and formed their stars earlier and more rapidly than low mass galaxies, 
in agreement with \citet{Gallazzi2006}. Therefore, the downsizing  pattern \citep{Cowie1996} could be applied to the assembly 
process of ellipticals at $z<1$. The downsizing trend is also confirmed by a recent work based on a spectroscopic sample up to $z \sim 1$ 
in the COSMOS field \citep{Pozzetti2010} that showed how the ETGs increase in number density with cosmic time faster for decreasing M.
 They also found that the number density of blue or spiral galaxies with M $> 10^{10}$M$_{\odot}$ remains almost constant, 
while the most extreme population of star-forming galaxies at intermediate/high mass is rapidly decreasing in number density with cosmic time. 
The authors suggest a transformation from blue active spiral galaxies of 
intermediate mass into blue quiescent and successively into red passive types with low specific star formation. This is in agreement with
the result  obtained by \citet{Faber2007} who concluded that most present-day E/S0 near $L^{*}$ arose from blue galaxies with ongoing star-formation 
that were quenched after $z\sim1$ and then migrated to the red sequence. The properties of nearby ETGs support a mixed scenario in which
quenched galaxies enter the red sequence via wet, gas-rich mergers, followed  by a limited number of dry, stellar mergers along the sequence.
The evolution of the stellar mass function of the ETGs up to $z \sim 1$ has also been studied by other authors ( e.g., \citealt{Bell2004}, 
\citealt{Brown2007},\citealt{Taylor2009}, \citealt{Nichol2007}), finding a moderate increase in the number density of quiescent galaxies 
with cosmic time which is mass dependent.
 
 At higher redshifts \citet{Arnouts2007} and \citet{Cirasuolo2007} found a rapid
rise in the space density of massive red sequence galaxies from $z \thicksim$ 2 to
$z \thicksim$ 1. These results are confirmed by \citet{Ilbert2010} who found that
the stellar mass density of quiescent galaxies of all masses increases by 1.1 dex between
 $z \sim 1.7$ and  $z \sim 1$, while it evolves only by 0.3 dex between
$0.8-1$ and  $z \sim 0.1$. This trend is also confirmed by \citet{Fontana2009}, who found that 
the fraction of quiescent galaxies increases from $15-20 \%$ at  $z > $ 2 to $\sim 40\%$ at $z \sim 1.2$,  
and recently by \citet{Nicol2011}, who found that the red sequence massive galaxies ($M > 10^{11}$ M$_{\odot}$) increase in
mass density by a factor $\sim 4$  from $z \sim 2$ to 1.

Some groups have claimed the existence of a substantial population of massive 
galaxies at $ z > 4$ (e.g., \citealt{Mobasher2004}, \citealt{Yan2006}, \citealt{Rodighiero2007}, \citealt{Fontana2009}, \citealt{Mancini2009}).
\citet{Marchesini2009} and \citet{Marchesini2010} found that the number density of most massive galaxies ($M > 3 \times 10^{11}$ M$_{\odot}$)
seems to evolve very little from $z \sim 4$ to $z \sim 1.5$, with a larger subsequent evolution down to $z \sim 0.1$. 
These results are broadly consistent with the work from \citet{vanDokkum2010}, where the authors derived an increase of massive galaxies
by a factor of $\sim 2$ since $z \sim 2$ to the present. They concluded that this growth is likely dominated by mergers,
 as in situ star-formation can only account for $\sim 20\%$ of the mass build up from $z=2$ to $z=0$.

Several studies suggest that the critical redshift range where the strongest evolution
and assembly took place is 1 $\leqslant  z \leqslant 2$ (eg., \citealt{Arnouts2007}, \citealt{Abraham2007}).
However, while the results up to $z \sim1$ are quite solid thanks to the
large samples of galaxies and to the spectroscopic infomation available, the samples of ETGs studied up to now  at $z > 1.4$ are still small.
Besides, ETG are the more clustered galaxies in the universe, making their study very complicated given field to field variations which introduce 
large uncertainties due to the cosmic variance. Anyway, many works have identified these objects spectroscopically, showing that passive/quiescent 
galaxies with elliptical morphology may exist up to $z \sim 2.5$ (\citealt{Cimatti2004b}, \citealt{McCarthy2004}, \citealt{Cimatti2008}). 
There are even some candidates of ETGs at $z > 4-5$, but their nature is still unknown as they are optically too faint for being
 spectroscopically observed.
Therefore, the uncertainties involved in the study of the evolution of quiescent galaxies at high redshifts are still large due to the 
observational difficulty to identify large samples of ETGs at high $z$ ($ > 1.5$). In particular, long
wavelength data are necessary to cover the optical-near infrared part of the
spectrum of high redshift galaxies, a spectral range fundamental when
calculating the galaxy stellar masses. In addition, the collection of a large
sample of high redshift galaxies requires a combination of large areas and deep
near-IR (NIR) observations. Due to these difficulties, the study of the galaxy stellar mass function (GSMF hereafter) divided 
by galaxy types at $z >1.5$  has  been explored only roughly and our knowledge of the evolution of galaxies at
high $z$ is still rather poor.

Motivated by the limited information available at  $z \geq 1.5 $, we present in
this paper the evolution of the GSMF with cosmic time at $z \geq 1.4 $ for the
total galaxy population and for the sample divided on the basis of the galaxy star-forming activity in the
COSMOS field. The COSMOS survey is the perfect wide area sample to look for rare and clustered
objects such as the old and massive ETGs with much improved statistics. This
survey \citep{Scoville2007} presents many advantages when compared with other
surveys. First of all, the continuous coverage of $\sim 2$ deg$^{2}$ substantially reduces the
effect of cosmic variance. Besides, its extensive multi-$\lambda$ coverage
allows to calculate accurate photometric redshifts \citep{Ilbert2009} and to
study the sources based on their spectral energy distributions (SEDs), which
is a more physical and complete approach than using only colors or morphological
information. Secondly, the multiwavelength data include deep Spitzer/IRAC
($3.6-8.0 \mu$m; Sanders et al. 2007) and \textit{$K_{s}$}-band data
\citep{McCracken2010}, crucial for estimating stellar masses at
$z\thicksim$2.

Galaxies of the present work are selected in the IRAC1 channel (3.6 $\mu$m), allowing to sample
the optical-NIR stellar bump at  $z >$ 1. 
In addition, data at red wavelength are crucial to detect galaxies that are
very faint in the optical (even undetected) but may contribute significantly  to,
or even dominate, the stellar mass density at high redshift (\citealt{Yan2000},
\citealt{Franx2003}, \citealt{Rodighiero2007}). The extensive multi-band data are used to separate galaxies
into different populations based on their SEDs. The present work extends to higher redshifts the study
of \citet{Ilbert2010}. The sample of  \citet{Ilbert2010} was selected at 3.6 $\mu$m as the present one,
but their restriction to IRAC sources with an  \textit{$i^{+}$}-band counterpart ($i^{+} <$ 25.5) limited their study
to  $z \lesssim 2$ and to larger stellar masses. Our work was possible after a careful analysis of sources not detected in the optical.

The paper is organized as follows: in Section 2 we introduce the COSMOS data and
the cross-correlation method. In Sections 3 and 4 we describe the method to
estimate the photometric redshifts and the galaxy stellar masses, while the
criteria used to split the sample in different populations is described in
Section 5. In Section 6  we study the main properties of the high redshift quiescent galaxies
and in Section 7 we present the GSMF for the whole sample and for the
different populations, discussing our results in Section 8. Finally, in Section
9 we summarize our conclusions. 

Throughout this paper we use standard cosmology ($\Omega_{m}=0.3,
\Omega_{\Lambda}=0.7$), with $H_{0}=70$  km s$^{-1}$ Mpc$ ^{-1}$. Magnitudes are
given in AB system. The stellar masses are given in units of solar
masses (M$_{\odot}$) for a Chabrier \citep{Chabrier2003} initial mass function
(hereafter IMF). The stellar masses based on a Salpeter IMF were converted into
a Chabrier IMF by adding $-0.24$ dex to the logarithm of stellar masses
(\citealt{Arnouts2007}) .

\section{Data}

We have built a multi-wavelength IRAC 3.6 $\mu$m  selected catalogue in the
COSMOS field by associating  the IRAC sources with the optical and MIPS
counterparts through the likelihood ratio method (see Section 2.2). For sources
undetected at $5\sigma$ in the \textit{$i^{+}$}-band selected catalogue we have performed a nearest match with the \textit{$K_{s}$}-selected
catalogue. The IRAC catalogue is 95$\%$ complete at 5.75 $\mu$Jy (22.0
mag).
The final catalogue consists of 78649 sources with $mag_{3.6 \mu m} < 22.0$:
$95\%$ (74742) of them with an optical counterpart, $4.5\%$ (3554) with no optical
counterpart but a \textit{$K_{s}$}-band counterpart  and $0.5\%$ (353) with only IRAC
detection. Besides, $14\%$ (11352)  of the IRAC sample are also detected in the MIPS band.

\subsection{Catalogues}

\begin{enumerate}
 \item 
\textit{IRAC}

The COSMOS IRAC catalogue \citep{Sanders2007} includes photometry in the 4 IRAC
channels (3.6, 4.5, 5.8, 8.0 $\mu$m ) for sources that have a measured flux in
IRAC Channel 1 above 1 $\mu$Jy (23.9 mag) in the COSMOS field. We used the $1.9"$
aperture corrected flux and we considered sources with $mag_{3.6 \mu m} <$ 22.0.
 
\item
\textit{Optical}

The optical catalogue \citep{Capak2007} is an \textit{$i^{+}$}-band ($i < 26.5$ mag) selected catalogue
in the area of the Subaru survey ($2$ deg$^{2}$) and includes photometry in 9 bands
(\textit{$u^{\ast}$, $ B_{J}$, $ g^{+}$, $ V_{J}$,$ r^{+}$, $ i^{+}$, $z^{+}$, $J$,
$K_{s}$}). We applied systematic offsets derived to optimise the photometric redshifts by \citet{Ilbert2009} and auto-offset
corrections (which account for the aperture correction) to the optical catalogue.

\item
\textit{24 $\mu$m}

The 24 $\mu$m catalogue \citep{LeFloch2009} is composed by sources detected
above 0.15 mJy (18.5 mag).

\item
\textit{$K_{s}$}

The \textit{$K_{s}$}-selected catalogue from \citet{McCracken2010} contains galaxies detected at
 \textit{$K_{s}$} and additional data from $B_{J},i^{+},z^{+}$ and $J$ bands. The catalogue is complete down to $K_{s} \sim 23$ mag .
We applied a magnitude dependent aperture correction to the fluxes, as explained in \citet{McCracken2010}.

\end{enumerate}

\subsection{The cross-correlation}

We have performed a cross-correlation of the IRAC sources with the optical and
MIPS catalogues through the likelihood ratio method.

 The likelihood ratio technique (\citealt{Sutherland1992};
\citealt{Ciliegi2003}) has the advantage of taking into account not only the
position (as for the nearest neighbour method, commonly used in literature), but
also the magnitude distribution of the background sources, helping to reduce
wrong identifications. The likelihood ratio LR is defined as the ratio between
the probability that the source is the correct identification  and the
corresponding probability  of a background, unrelated object:

\begin{equation}
LR=q(m)f(r)/n(m)
\end{equation}

where $q(m)$ is the expected probability distribution, as a function of
magnitude, of the true counterparts, $f(r)$ is the probability distribution
function of the positional errors of the IRAC sources
(assumed to be a two-dimensional Gaussian) and $n(m)$ is the surface density of
background objects with magnitude $m$.
In order to derive an estimate for $q(m)$ we have first counted all objects in
the optical/MIPS catalogue within a fixed radius around each source
$(total(m)$). This has then been background subtracted and normalized to
construct the distribution function of real identifications:

\begin{equation}
q(m)=\frac{real(m)Q}{\sum_{i}real(m)_{i}}
\end{equation}

where $real(m)=total(m)-n(m)$ (the sum is over all the magnitude bins of the
distribution) and Q is the probability that the optical/MIPS counterpart of the
source is brighter than the magnitude limit of the optical/MIPS catalogue
($Q=\int^{m_{lim}} q(m)dm$). 
Finally, a best threshold value for LR $(LR_{th})$ is chosen in order to
maximize both the completeness (avoid missing real identifications) and
reliability (keep spurious identifications as low as possible) of the
associations.

\subsubsection{Results}

In this Section we present the final multi-wavelength catalogue created by
cross-correlating the IRAC catalogue with the optical and the MIPS catalogues.
For the sources with no $5\sigma$ \textit{$i^{+}$}-band counterpart we have performed a match with the
 \textit{$K_{s}$}-band catalogue.

\begin{enumerate}
 \item 
\textit{IRAC + Optical}

We considered sources in the optical catalogue with flux in the \textit{$i^{+}$}-band  $> 5\sigma$,
and eliminated the masked areas and the stars from both catalogues. We performed
a match through the likelihood ratio method considering a maximum radius of 3
arcsec (in order to maximize the overdensity around the IRAC position) and a
dispersion of $\sigma=1 "$ for the distribution of the positional errors.
The distribution of the \textit{background}, the \textit{total} and the \textit{real} counts is shown in
Fig. 1. 

\begin{figure}
\centering
\includegraphics[angle=90,width=0.5\textwidth]{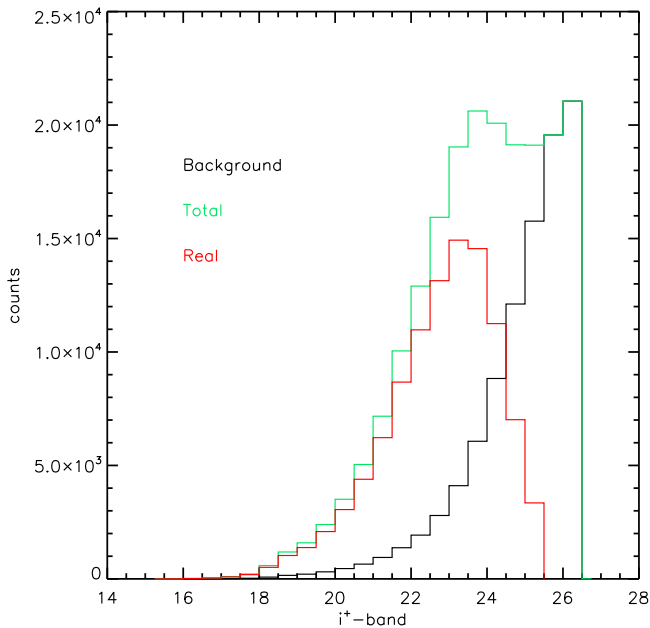}
\caption{Magnitude distribution of the background (black line), total (green
line) and real counts (red line) in the $i^{+}$-band.}
\end{figure}

The majority ($96\%$) of the associations have a separation $\leq$ 1 arcsec, 
with the median separation being $\sim$ 0.34 arcsec.
For a non negligible fraction of the associations with a separation $>$ 1 arcsec,
the photometric redshift we obtained (see Section 3) was not in agreement with the spectroscopic one,
meaning a probable spurious IRAC-optical association. Therefore,  we have considered as reliable
only matches within a separation of 1 arcsec, which is a standard separation used in literature when
cross-correlating optical and IRAC catalogues.
 To reduce contamination from Active Galactic Nuclei (AGN) we have also
eliminated the IRAC sources with an optical counterpart asociated to an X-ray detection with XMM-Newton ($1.5 \%$ of the IRAC sample).
Finally, we obtained a reliable optical counterpart for 74742 IRAC sources ($95\%$).
 We will refer to these sources as the "optical sub-sample"  hereafter.

\item
\textit{IRAC+24 $\mu$m}

We found 11352 IRAC sources (14$\%$) with a reliable MIPS counterpart detected at  $5\sigma$. The likelihood ratio method, as described above,
has been used.

\item
\textit{IRAC+$K_{s}$}

For the sources without an optical counterpart we have performed a match with the \textit{$K_{s}$}-selected catalogue.
In this case we used the nearest neighbour technique with a maximum separation
of 1 arcsec. The similar wavelengths of the \textit{$K_{s}$} and IRAC1 bands allows to
perform this match with a low probability of misidentification. $90\%$ (3554 sources over 3907) of the IRAC
sources with no optical counterpart, have a \textit{$K_{s}$}
counterpart ( "\textit{$K_{s}$} sub-sample" hereafter). As already mentioned, the \textit{$K_{s}$}-band catalogue
from  \citet{McCracken2010} contains also information in additional bands ($B_{J},i^{+},z^{+}$ and $J$).
The median magnitude  in the $i^{+}$-band of these sources is $i^{+} \sim 26$, i.e., these sources are optically very faint,
consistent with them being not in the  "optical sub-sample ".

\end{enumerate}

Finally, we have 353 sources for which we did not find neither an \textit{$i^{+}$}-band counterpart, nor a \textit{$K_{s}$} counterpart and for which we have only the 
IRAC bands information  ( "IRAC sub-sample" hereafter) and, eventually, the 24 $\mu$m (50 sources). The results of the cross-correlation and
the final numbers of our catalogue are summarized in Table 1.

\begin{table}
\begin{tabular}{|r|r|r|}
\hline
 & N ($\%$ of the whole sample) & MIPS detection \\
\hline
IRAC sample (all) & 78649 & 11352 \\
\hline
Opt. sub-sample & 74742 ($95\%$) & 10779 \\
\textit{$K_{s}$} sub-sample &  3554 ($4.5\%$)& 523 \\
IRAC sub-sample & 353 ($ 0.5\%$) & 50 \\
\hline
\end{tabular}
\caption{Number of sources detected in the optical, \textit{$K_{s}$} and MIPS catalogues.}
\end{table}

\section{Photometric Redshift}

The only efficient way to estimate the redshifts of the galaxies for the whole sample, given the large
number of sources considered, is to determine their photometric redshifts through
a detailed SED-fitting procedure. 

Photometric redshifts were derived using the  \textit{Le Phare} code
(\citealt{Arnouts2001}, \citealt{Ilbert2006}), which computes photometric
redshifts based on a simple $\chi^{2}$  template-fitting procedure. We used the
COSMOS SED library provided by \citet{Ilbert2009}, which contains a set of templates
composed of  ellipticals and  spirals from the \citet{Polletta2007} library and
also includes blue galaxies from the \citet{BC2003} population synthesis models.
We added 6 simple stellar population (SSP) models of different ages
(0.05, 0.5, 1.0, 1.5, 2.0 and 3.0 Gyr) from \citet{Maraston2005} to account for
passively evolving galaxies at high redshifts. The final library is composed of
7 ellipticals, 12 spirals, 12 starbursts and 6 SSP templates. Some representative templates
of the library are shown in Fig. 2. This set of templates maximizes the accuracy
of the derived photometric redshifts ($z_{p}$) when compared with the available spectroscopic redshifts ($z_{s}$)
(see Section 3.1.).

\begin{figure}
\centering
\includegraphics[angle=0,
width=0.5\textwidth]{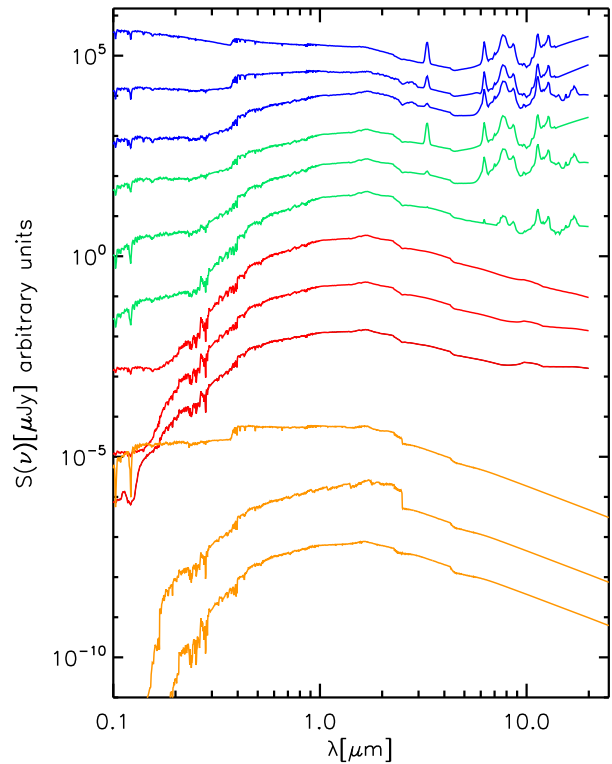}
\caption{Representative templates of the SED  library. Orange: simple stellar
population models of 0.05, 1.0 and 3.0 Gyr \citep{Maraston2005}. Red: elliptical \citep{Polletta2007}.
Green: spirals \citep{Polletta2007}. Blue: blue galaxies \citep{BC2003}. 
}
\end{figure}

We allowed  3 different extinction laws (\citealt{Calzetti2000}, Calzetti
modified and \citealt{Prevot1984}) to be applied to each template. The Calzetti 
modified extinction law includes a UV bump at 2175 $\AA{}$ (\citealt{Ilbert2009}),
 which is fundamental to have good $z_{p}$ at  $z > 1.5$. We used 9 different values of E(B-V): 0, 0.05, 0.1, 
0.15, 0.2, 0.25, 0.3, 0.4 and 0.5.

We set an upper limit  (5$\sigma$) to each band when there is no detection, as shown in Table 2.
This 5$\sigma$ limit has been chosen following a conservative approach, i.e.,
as the brightest magnitude value of the sources in the catalogue detected with S/N=5.
 We fitted our data from the $U$ to the 24$\mu$m band.
The $\chi^{2}$ minimization procedure is highly affected by the adopted photometric errors 
(see \citealt{Ilbert2009} for a detailed discussion of the issue). We have
increased the flux error  by 6, 5 and 8$\%$ for the optical, IRAC and MIPS bands
respectively. 

\begin{table}
\begin{tabular}{|c|c|c|}
\hline
band & optical sub-sample & \textit{$K_{s}$} sub-sample\\
\hline
$u^{*}$ & 25.3 & -\\
$b_{J}$ & 25.7 & 27.0\\
$g^{+}$ & 25.5 & -\\
$V_{J}$ & 25.5 & -\\
$r^{+}$ & 25.6 & -\\
$i^{+}$ & 25.1 & 26.2\\
$z^{+}$ & 24.5 & 25.0\\
$J$ & 22.0 & 22.0\\
$K$ & 22.0 & -\\
$IRAC1$ & - & -\\
$IRAC2$ & 23.3 & -\\
$IRAC3$ & 21.2 & -\\
$IRAC4$ & 21.0 & -\\
$MIPS$ & 18.5 & -\\

\hline
\end{tabular} 
\caption{5$\sigma$ limit adopted for each band}
\end{table}

\subsection{Photometric Redshift accuracy}

We were able to test the qualitity of our derived photometric redshifts for the
IRAC sources with \textit{$i^{+}$}-band detection by comparison with a large sample of
spectroscopic redshifts. In particular, we made use of 8176 sources with derived
spectroscopic redshift at a very high confidence level  ($99.8 \%$, flags 3.1, 3.5, 4.1, 4.5) from the
zCOSMOS faint and bright surveys (\citealt{Lilly2007, Lilly2009}).
We estimated the redshift accuracy as $\sigma_{\Delta z/(1+z_{s})}$ (where
$\Delta z= zp-zs$ is the difference between the photometric and the spectroscopic
redshifts) using the normalized median absolute deviation defined as 
$1.48\langle\vert z_{p}-z_{s}\vert/(1+z_{s})\rangle$. We also defined the percentage of
catastrophic errors, $\eta$, as the objects with $\vert
z_{p}-z_{s}\vert/(1+z_{s}) > 0.15$.
We obtained an accuracy of $\sigma_{\Delta z/(1+z_{s})}=0.060$ and a percentage
of catastrophic failures of $\eta=3.3\%$ as can be seen in Fig. 3.
As a comparison, \citet{Ilbert2009} using the bright ($i^{+} < 22.5$ mag) sample
with 30 bands and narrow filters obtained  $\sigma=0.007$ and  $ < 1 \%$
of catastrophic errors for their derived photometric redshifts, while the average
value of $\sigma$ obtained by  \citet{PG2008} using an IRAC selected catalogue
with data from the UV to the MIR is 0.055. Given the fact that we use 14
broad bands, we find our result comparable to those in literature.

\begin{figure}
\centering
\includegraphics[angle=0,
width=0.5\textwidth]{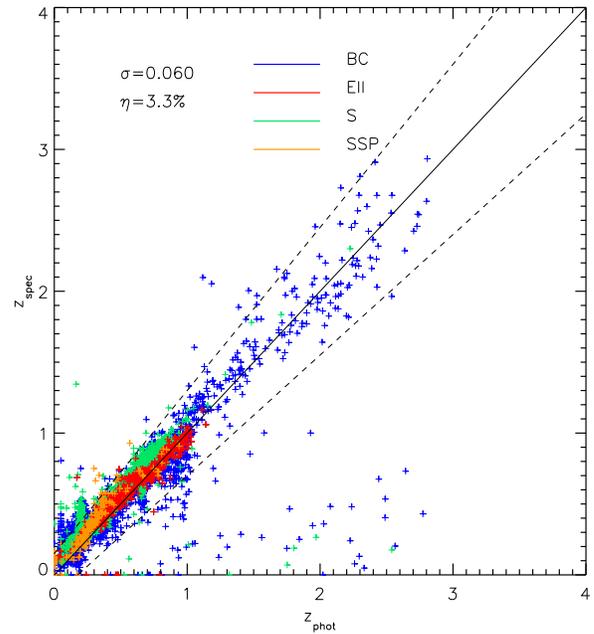}
\caption{Comparison between $z_{p}$ and $z_{s}$ for the 8176 sources with
optical and IRAC detection of the zCOSMOS sample. The value of $\sigma$ and
$\eta$ are shown in the  plot. The different colors represent sources fitted by
different SED templates as explained in Fig. 2. Dashed lines represent  $\vert
z_{p}-z_{s}\vert/(1+z_{s}) = 0.15$, i.e., sources out of that region are considered as catastrophic errors. }
\end{figure}

We also tested the reliability of our $z_{p}$ for the faintest sources,
with $mag_{3.6 \mu m} > 21.0$ and $i^{+} > 22.5$, as shown in Figs. 4 and 5. 
The value of $\sigma$ for the faint  $i^{+}$-band sources ($ \sigma=0.057$) is comparable to
 the value of $\sigma$ for the whole sample, while the number of catastrophic errors increases ($\eta=12\%$). For the faint IRAC1 sample both $\sigma$ and $\eta$
are smaller than for the whole sample ($\sigma=0.037$, $\eta=1.9\%$).  
This may surprise the reader, but it must be noticed taht the  values for the whole spectroscopic sample are highly affected
by a significant dispersion on low redshift sources ( $z_{s} \sim$   $0.3-0.5$, see Fig. 3), which are present in a 
smaller fraction in the faint IRAC sample. 

\begin{figure}
\centering
\includegraphics[angle=0,
width=0.5\textwidth]{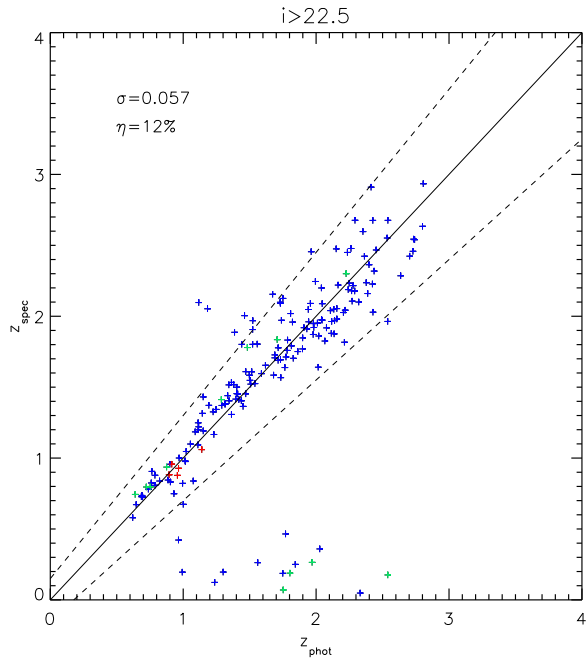}
\caption{Comparison between $z_{p}$ and $z_{s}$ for 183 sources with $ i^{+}$-band $> 22.5$. Colors and dashed lines as in Fig. 2.}
\end{figure}

\begin{figure}
\centering
\includegraphics[angle=0,
width=0.5\textwidth]{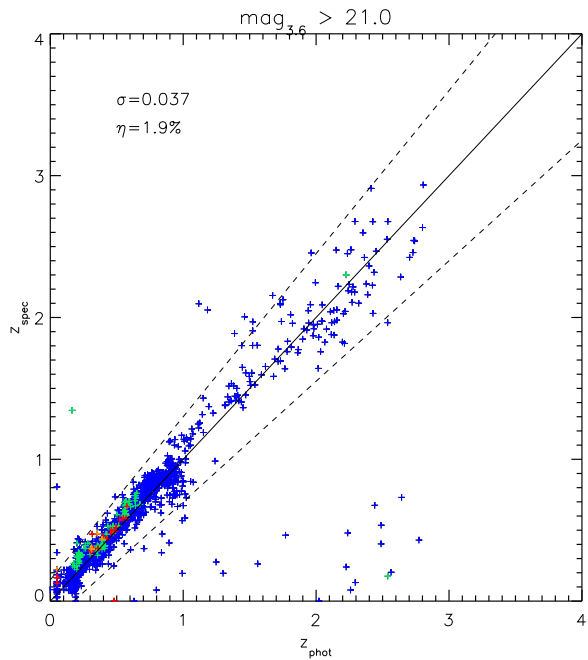}
\caption{Comparison between $z_{p}$ and $z_{s}$ for 2381 sources with  $mag_{3.6 \mu m} >
21.0$. Colors and dashed lines as in Fig. 2.}
\end{figure}

\subsection{Photometric Redshift accuracy for the IRAC sources with no I band counterparts}

For the sources with no \textit{$i^{+}$}-band counterpart it is not possible to directly
compare the photometric redshifts with spectroscopic redshifts, since no $z_{s}$ 
are available for these sources. To test the quality of the derived photometric
redshifts we have made use of the $1\sigma$ dispersion of the $z_{p}$
distribution, given as an output of the \textit{Le Phare} code. The mean
error on the $z_{p}$ for the whole sample is $\sigma_{z}=0.17$, which corresponds to 
 $\sigma_{\Delta z/(1+z_{s})} \sim 0.057$ at $z \sim 2$, meaning that  the $1\sigma$ dispersion of the $z_{p}$
distribution is consistent with the $\sigma_{\Delta z/(1+z_{s})} $ derived from the $z_{p}$ and $z_{s}$ comparison.
The redshift uncertainties derived from \textit{Le Phare} have been used in the Montecarlo simulations (see Section 7).

In Figs. 6,7 and 8 we show some examples of SED fitting for sources fitted by
different templates (elliptical (Ell), simple stellar populations (SSP), spirals
(S) and blue galaxies or starbursts (SB)) for sources from the optical, \textit{$K_{s}$} and
IRAC sub-samples.

\begin{figure}
\centering
\includegraphics[angle=0,
width=0.5\textwidth]{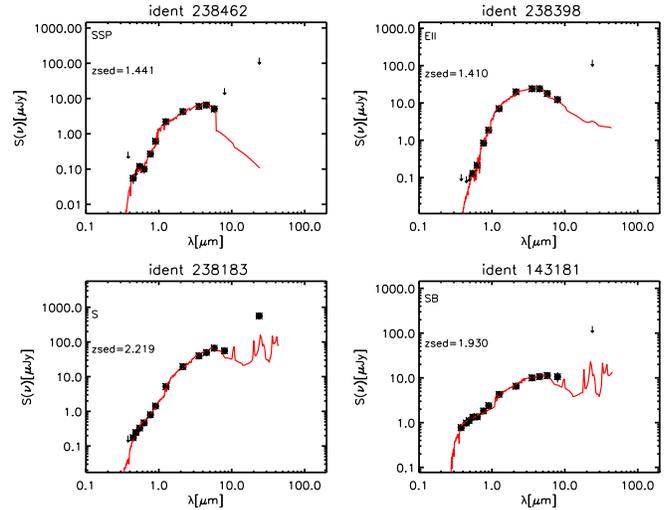}
\caption{Example of SED-fitting for 4 sources from the optical sub-sample. The
observed SED of each source (black filled circles) is shown with the
corresponding best-fit solution (red solid line), as well as the derived $z_{p}$
and best template model. Arrows represent upper limits (5$\sigma$) when no
detection is found in that band.}
\end{figure}

\begin{figure}
\centering
\includegraphics[angle=0,
width=0.5\textwidth]{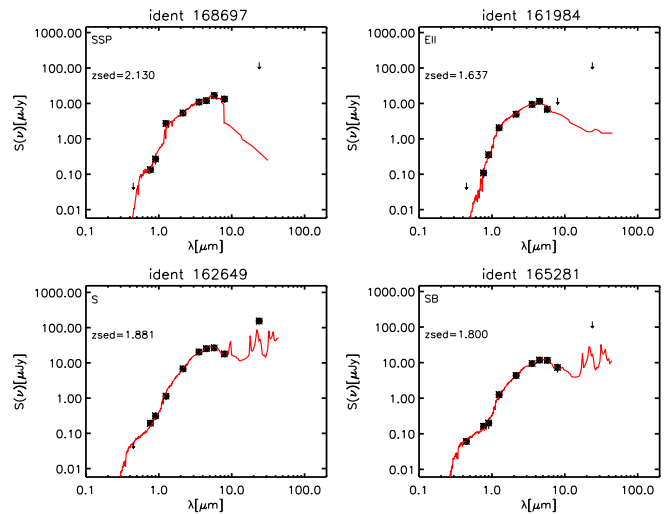}
\caption{Example of SED-fitting for 4 sources from the \textit{$K_{s}$} sub-sample, i.e.,
sources not present in the \textit{$i^{+}$}-band catalogue (\citealt{Ilbert2010}) but with a \textit{$K_{s}$} counterpart (\citealt{McCracken2010}).}
\end{figure}

\begin{figure}
\centering
\includegraphics[angle=0,
width=0.5\textwidth]{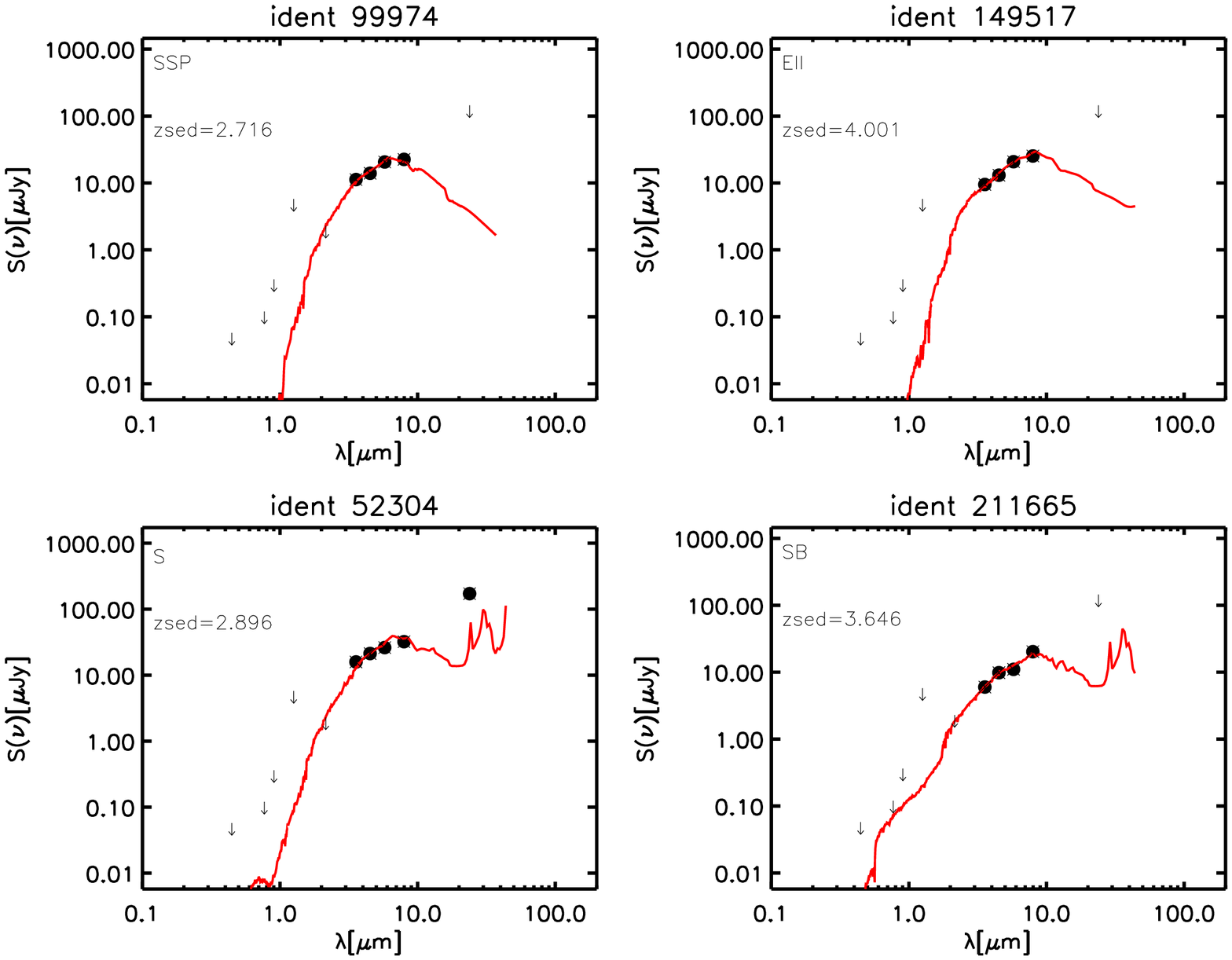}
\caption{Example of SED-fitting for 4 sources from the IRAC sub-sample, i.e.,
sources only detected in the IRAC bands.}
\end{figure}

\subsection{Redshift distribution}

 Fig. 9  shows the redshift distribution for each sub-sample, highlighting the contribution from different
 classes of best-fitting templates. Templates have been divided in 3 classes: elliptical/old SSP (older than 1 Gyr), 
blue/young SSP (younger than 1 Gyr) and spirals.
It can be observed that these redshift distributions are not smooth, but show some peaks at given redshifts, e.g., at $z \sim$ 1.0, 2.0 
for the \textit{$K_{s}$} sub-sample or a very prominent peak at z $\sim$ 4.0 for the IRAC sub-sample. Note that the peak at $z\sim 4$ in the
IRAC sub-sample is not used in this paper since we limit our  analysis to $z < 3.0$ and also the small number of sources involved in this peak 
($< 0.2 \%$ of the sample ). This kind of features in the redshift distribution
 are presumably not real, but due to the determination of the redshifts through a SED fitting procedure. Spiky redshift distributions
have also been reported in literature, e.g., by \citet{PG2008} 
and \citet{Franx2008} among others.
The number of sources in each subsample with a reliable $z_{p}$ is shown in Table 3, 
as well as the number of sources for each sub-sample fitted with each class of template. 
The number of high redshift ($z \gtrsim 1.4$ and  $z \gtrsim 2.5$) sources for each sub-sample is also shown.

\begin{figure}
\centering
\includegraphics[angle=90,
width=0.5\textwidth]{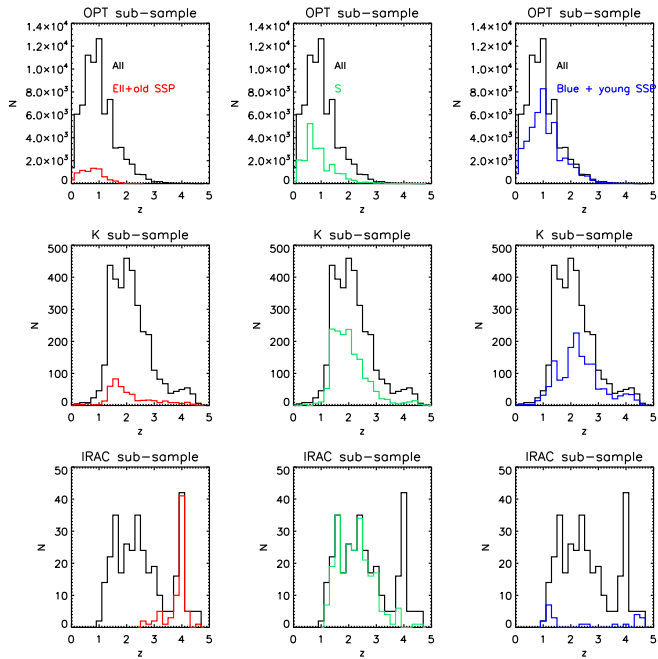}
\caption{Redshift distribution for the different sub-samples. In the 3 upper panels we
 show the redshift distribution of the optical sub-sample (black line), as well as the redshift distribution of the sources fitted by
an elliptical/old SSP template (red line), a spiral template (green line) or a blue/young
 SSP template (blue line). The same is plotted for the \textit{$K_{s}$} and the IRAC sub-samples in the middle 
and bottom panels, respectively. }
\end{figure}

\begin{table}
\begin{tabular}{|c|c|c|c|}
\hline
   & Opt. sub-sample & \textit{$K_{s}$} sub-sample & IRAC sub-sample\\
\hline
All             & 74742  & 3554   & 328\\

Ell/old SSP     & $10\%$ & $12\%$ & $22\%$\\
S               & $28\%$ & $45\%$ & $71\%$\\
Blue/young SSP  & $62\%$ & $43\%$ & $7\%$\\
\hline
$z \geq 1.4$       & $21\%$& $88\%$ & $91\%$\\
$z \geq 2.5$       & $2.3\%$& $27\%$ & $47\%$\\
\hline
\end{tabular} 
\caption{Percentages of best fit templates and high redshift sources ($z \geq
1.4$ and $z \geq 2.5$) for the different sub-samples.}
\end{table}

It is clear how differently the three sub-samples behave, with $88\%$
and $91\%$ of the \textit{$K_{s}$} and the IRAC sub-samples respectively found at high redshift
($z \geq 1.4$), in comparison to $21 \%$ of the optical sub-sample. For very high redshift sources ($z \gtrsim 2.5$)
 the differences are even more significant, since only 2.3 $\%$ of the optical sub-sample is found at those redshifts, while
this percentage is $ 27\%$ and $47\%$ for the \textit{$K_{s}$} and IRAC sub-samples, respectively.
Also the relative fractions of best-fitting templates are very different.
 On the one hand, $10\%$ of the optical sub-sample is fitted by an elliptical/old SSP model, while
for the IRAC sub-sample the fraction is double ($22\%$). On the other hand,
only $7\%$ of the IRAC sub-sample is fitted by a blue/young SSP model, while these templates
represent  $62\%$ of the solutions for the optical sub-sample. These
percentages are indicative of the importance of an IRAC selected sample to study
the properties of high redshift galaxies whose SED is reproduced by quiescent populations.

\section{Galaxy stellar masses}

To study the evolution of the galaxy  stellar mass  function of sources at high
redshift we have selected the sources with $z \geq 1.4$. When available, we have
made use of  $z_{s}$ (654 out of 19042 sources, $ \sim 3.4\%$ of the sample).
\newline
The galaxy stellar masses have been derived by means of the  $\textit{Lephare}$
 code by fitting our data (up to 5.8$\mu$m) with a set of SED templates from
\citet{Maraston2005} with star formation histories exponentially declining with time as
$SFR \propto e^{-t/\tau}$. We used 9 different values of $\tau$ (0.1, 0.3, 1.0,
2.0, 3.0, 5.0, 10.0, 15.0 and 30.0 Gyr) with 221 steps in age. The metallicity
is solar and the IMF Chabrier. Dust extinction was applied using the
\citet{Calzetti2000} extinction law, with a maximum E(B-V) value of 0.5.
 We imposed to the derived age of the
galaxies to be less than the age of the Universe at that redshift and  greater
than $10^{8}$ years (the latter requirement avoids having galaxies with extremely
high specific star formation rates, $SSFR=SFR/M$ (SSFR hereafter). 
The \citet{Maraston2005} models include a better treatment of the thermally pulsing
asymptotic giant branch (TP-AGB) phase, which has a high impact on modelling the
templates at ages in the range $0.3 \lesssim  t \lesssim 2$ Gyr, where the fuel
consumption in this phase is maximum, specially for the near-IR part. Although some authors \citep{Kriek2010} have recently claimed
that the \citet{Maraston2005} models do not properly reproduce the observed data in a sample of post-starburst galaxies, i.e., at the time
when TP-AGB stars are thought to be most dominant, we have instead found good fits for our SEDs.
\citet{Pozzetti2010} measured an average systematic shift of 0.14 dex between the stellar masses
computed with the \citet{BC2003} models and the  \citet{Maraston2005}. We will
take into account this difference when comparing our results with those obtained with the  \citet{BC2003} models.

Uncertainties in the stellar mass derivation are due to a number of different assumptions in the
SED fitting procedure, like, for example, the use of different IMF, extinction laws, metallicities, star formation
histories (SFH) or SED template libraries. Tests on simulated catalogues
considering the effect on stellar mass estimates of different choices of
reddening law, SFHs,
metallicities and  SED libraries show a typical dispersion of the order of
$\sigma (logM) \sim 0.20$ (see \citealt{Bolzonella2009}), showing that the stellar
mass is a rather stable parameter in SED-fitting.

 The mean error on the stellar mass, as estimated from  the $1\sigma$ dispersion of the mass distribution (given as
an output of the \textit{Le Phare} code), is 0.16 dex, consistent with the uncertainties mentioned above.

\section{Galaxy classification}

We are interested in studying the characteristic evolution of the quiescent and
star-forming galaxies. To discriminate between these two types, we have made use of
a classification based on the SSFR derived
through the SED-fitting process. We considered 3 different populations:

\begin{enumerate}
 \item Actively star-forming galaxies: those for which $log (SSFR [$Gyr$^{-1}]) > -0.5$
\item Intermediate galaxies: those for which $-2 <log (SSFR [$Gyr$^{-1}]) < -0.5$.
\item Quiescent galaxies: those for which $log (SSFR [$Gyr$^{-1}]) < -2$.

We found 84 sources ($\sim 0.5 \%$ of the high redshift sample, i.e., sources with $1.4 \leq z \leq 3.0$) 
with  $log (SSFR [$Gyr$^{-1}]) < -2$ and detected in the  $24 \mu$m band. 
As the emission at 24 $\mu$m usually arises from dust heated by star-formation activity, we include these sources
 in the intermediate population.

\end{enumerate}

The limits on the SSFR values have been defined from the distribution of the SSFR values
(Fig. 10 for the optical sub-sample) and we have verified that they are consistent
with the best-fitting template classification.
 Most of the sources with very low SSFR are fitted
by an elliptical or old SSP  template, while those with the
highest activity are generally fitted by a young SSP or a
blue template, as should be expected on the basis of their colors. Considering the whole
sample, we have carefully studied the cases where star-forming sources are fitted
by an elliptical or old SSP template, as well as quiescent sources fitted by a
blue or a young SSP template. We found only 26 sources ($ \lesssim 0.1\%$ of the high redshift sample) fitted by
an elliptical or old SSP template with high activity. On the other hand, more than $94\%$ of the 639 sources ($3.5\%$ of the
high redshift sample) fitted by a blue template with low SSFR are not detected
in the \textit{U} band, which is a critical band to determine the star formation activity of a galaxy. 

In Table 4 we report the number of galaxies classified as star-forming, intermediate and quiescent,
 as well as the MIPS detected sources, in each redshift bin for sources with $1.4 \leq  z \leq 3.0$.


\begin{table}
\begin{tabular}{|r|r|r|r|r|r|}
\hline
     $z$    &all & quiescent  & intermediate &   star-forming & MIPS\\
\hline  
1.4-1.6   & 5142    & 515 & 1325  &3302 &261\\
1.6-2.0   & 6303    & 512 & 1445  &4346 &1286\\
2.0-2.5   & 4693    & 287 & 914   &3492 &1178\\
2.5-3.0   & 1794    & 77  & 415   &1302 &272\\
\hline
\end{tabular}
\caption{Number of galaxies with $1.4 \leq  z \leq 3.0$ in each redshift bin as classified by the SSFR.}

\end{table} 

\begin{figure}
\centering
\includegraphics[angle=0,
width=0.5\textwidth]{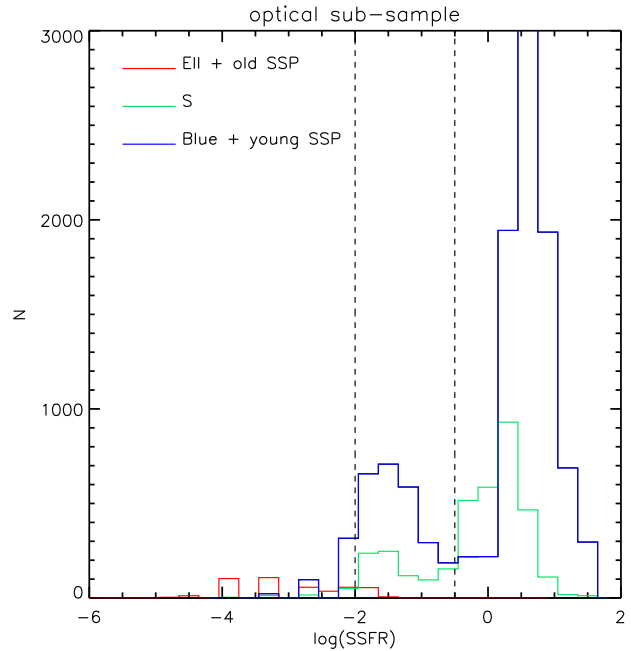}
\caption{SSFR distribution for the optical sub-sample of sources with $z \geq 1.4$. The colors represent the
different classes derived from the SED fitting templates. Dashed vertical lines represent the limits considered in SSFR to
divide sources as actively star-forming, intermediate or quiescent.}
\end{figure}

Other tests, such as the \textit{BzK} classification \citep{Daddi2004}, which separates
star forming and quiescent galaxies with respect to their $(B-z)$, $(z-K)$ colors
for sources with $1.4 \leq z \leq 2.5$, shows that galaxies for which low
activity is derived roughly occupy the passive region in the \textit{BzK} diagram 
($(z-K)-(B-z)< -0.2$ and $(z-K) > 2.5$), while for the star-forming galaxies,
their colors are consistent with the \textit{BzK} star-forming selection (Fig. 11). 
Only 3 sources defined as active star forming fall in the passive region of the \textit{BzK} diagram.

\begin{figure}
\centering
\includegraphics[angle=90,
width=0.5\textwidth]{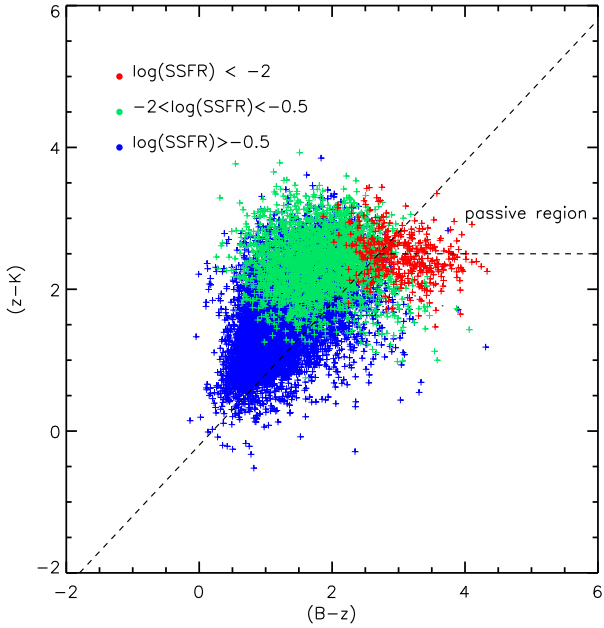}
\caption{\textit{BzK} diagram \citep{Daddi2004} for optically detected sources with $1.4
\leq z \leq 2.5$. The colors represent the different
classes derived from the SSFR criteria (red: quiescent, green: intermediate, blue: star-forming). The
dashed lines show the different regions of the \textit{BzK} diagram.}

\end{figure}

The lack of detection in some bands worsens the derived SSFR, thus complicating
a precise classification. We have derived mean SSFR errors making
use of the $1\sigma$ dispersion of the SSFR distribution, obtaining a mean SSFR error of  0.38 dex for the whole sample.

\section{ Properties of high-z quiescent galaxies}

\begin{figure}
\centering
\includegraphics[angle=0,
width=0.5\textwidth]{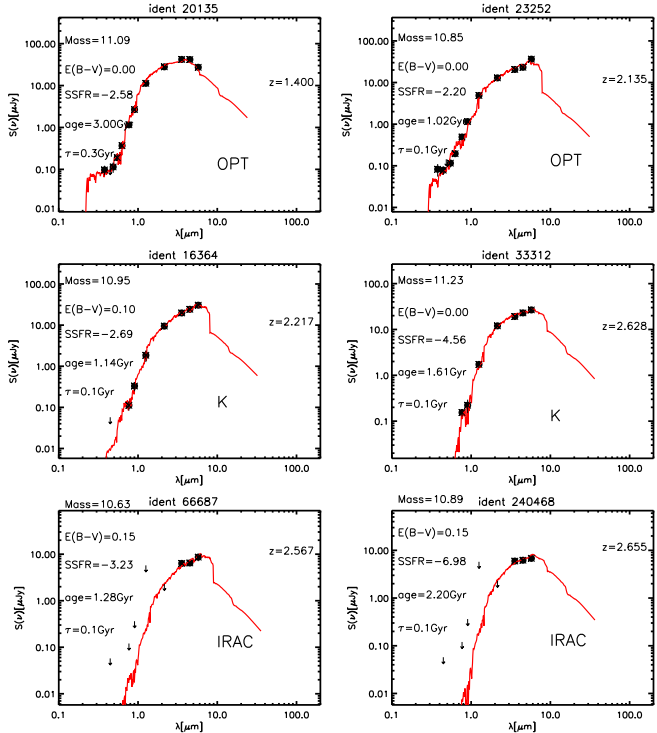}
\caption{SED-fitting of some high redshift quiescent galaxies. Also shown the main physical
parameters derived through the SED-fitting process (mass, extinction, SSFR, age, $\tau$, z).
The OPT, \textit{$K_{s}$} and IRAC labels indicate  the sub-sample at which each source belongs.}
\end{figure}

The evolution over the cosmic time of quiescent galaxies represents a key tool to understand 
the galaxy mass assembly. In this Section we summarize
the properties of the high redshift quiescent galaxies (also known as ETGs in literature) by analyzing the physical
parameters obtained through the SED-fitting procedure.

We find 1455 quiescent galaxies at $z \geq 1.4$. They are old and massive galaxies 
($\langle M \rangle \sim 10^{10.65}$M$_{\odot}$): more
than $56\%$ of them have $M\geq10^{10.6}$M$_{\odot}$ and $\sim 12\%$ have
$M\geq10^{11.0}$M$_{\odot}$. They have weakly star-forming stellar populations, with
ages ranging from 1 to 4.25 Gyr. Some of them are so old to reach the limit allowed by
the age of the Universe at their redshift, meaning very high formation redshifts.
They have \textit{e-folding} time scales $\tau \sim$ $0.1-0.3$ Gyr, i.e., they should
have formed in a very intense and brief starburst. They are characterized by very low dust
extinction (E(B-V) $\sim$ $0-0.15$), implying that their red optical colors are due to
evolved stellar populations and not to dust extinction. These
characteristics are in agreement with previous results (e.g., \citealt{Cimatti2009}, \citealt{Kriek2008}).

In Fig. 12 we show the SED-fitting and main physical parameters of 6 different
high redshift quiescent galaxies. We show examples from the optical, \textit{$K_{s}$} and IRAC sub-samples. 
We underline the importance of the upper limits in the optical and \textit{$K_{s}$} bands for the IRAC sub-sample
to constrain the SED model.

\section{The Galaxy Stellar Mass Functions}

\begin{figure}
\centering
\includegraphics[angle=90,
width=0.5\textwidth]{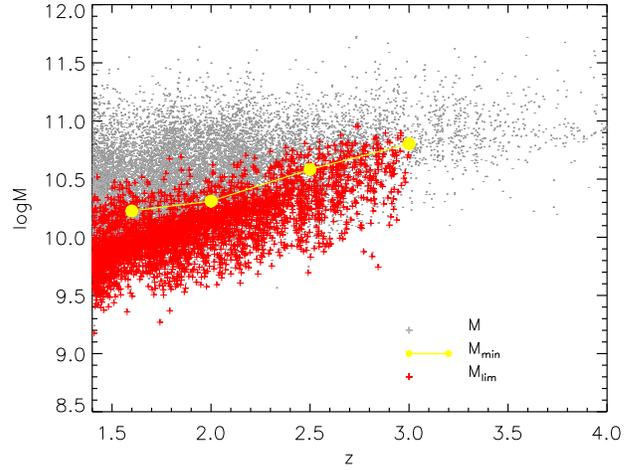}
\caption{Stellar mass as a function of redshift (small grey dots). Red crosses
are the $M_{lim}$ of the $20\%$ faintest galaxies at each redshift, while
$M_{min}(z)$ are plotted as yellow circles. }
\end{figure}

We study the evolution of the quiescent galaxies by deriving their mass function  
in four redshift bins. The 
mass function of the whole sample and of the star forming and intermediate galaxies 
are also estimated to compare the evolution of the different populations.
The classical non-parametric $1/V_{max}$ formalism \citep{Schmidt1968} has been used and 
data have been fitted to a Schechter function \citep{Schechter1976}. 
The galaxy stellar mass function GSMF is computed only in the non-masked regions with
a total covered area of 1.73 deg$^{2}$.

Estimating the stellar mass limit for a magnitude limited sample is not
straightforward due to the high range of possible M/L ratios for different galaxy
 populations and colors. To account for this effect we define, at each
redshift, a minimum mass, $M_{min}$, above which the derived GSMF is essentially
complete because all types of galaxies are potentially observable above this
mass.

Following \citet{Pozzetti2009}, we calculated for each galaxy its limiting stellar
mass, i.e., the mass it would have at its redshift if its observed magnitude 
were equal to the limiting magnitude of the sample ($mag_{3.6 lim}=22.0$);
$log(M_{lim})=log(M)+0.4(mag_{3.6}-mag_{3.6 lim}$). In order to derive a
representative limit for our sample we used the $M_{lim}$ of the $20\%$ faintest
galaxies at each redshift. We then defined the $M_{min}(z)$ as the upper
envelope of the $M_{lim}$ distribution below which lie $95\%$ of the
$M_{lim}$ values at each redshift. This $M_{min}$ corresponds to a 95$\%$
completeness limit at each redshift.
The distribution of mass with redshift, as well as the $M_{lim}$ and the
$M_{min}(z)$ calculated as explained above are shown in Fig. 13 .

\begin{figure*}
\centering
\includegraphics[angle=0,
width=1.0\textwidth]{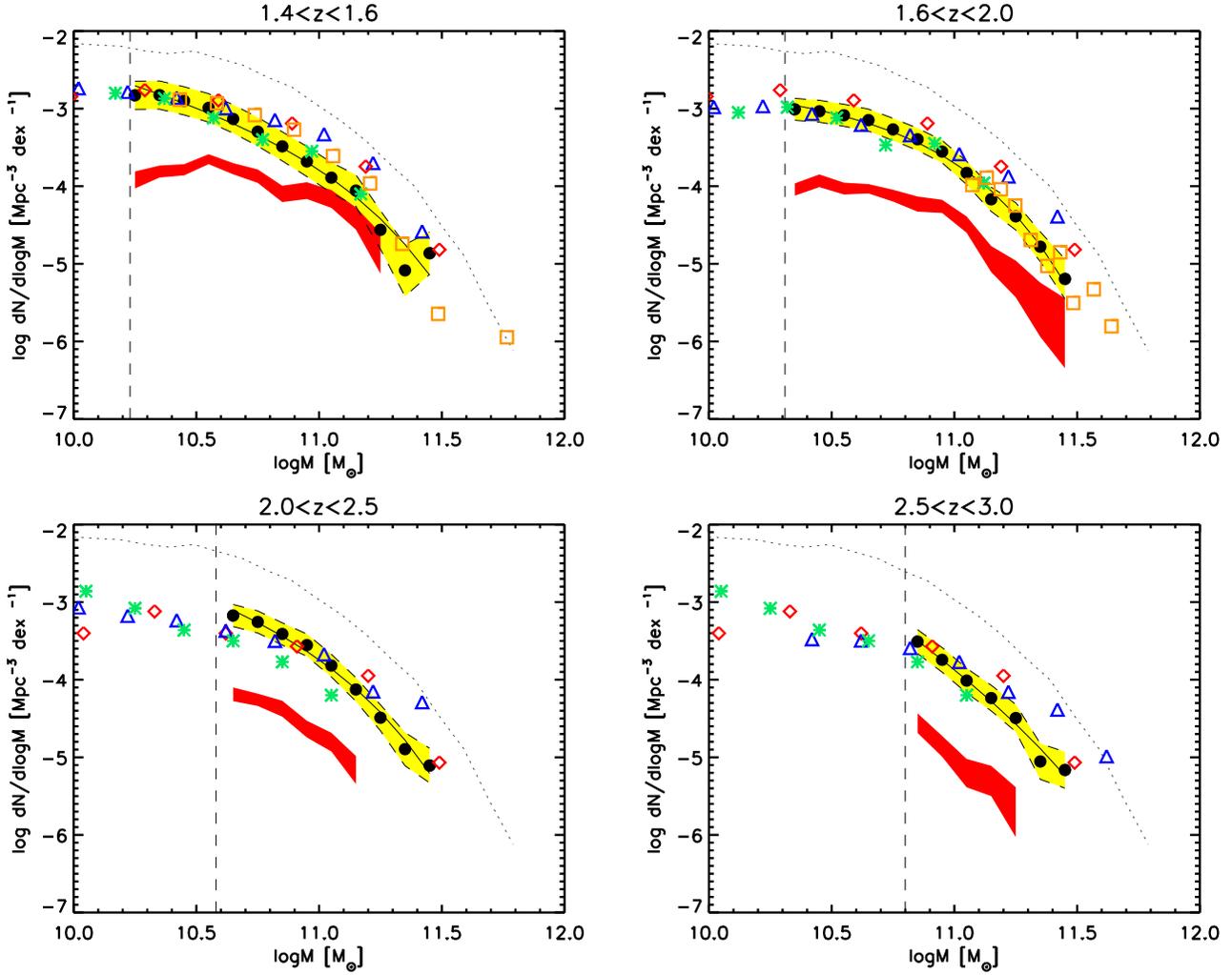}
\caption{Stellar mass function in 4 redshift intervals for the quiescent (red area) and the total (yellow area) sample.
 Our estimates at each redshift for the whole sample are plotted as filled black circles. Data are fitted with a Schechter
function (solid lines). The colored areas represent upper and lower limit of the
total and quiescent mass functions as derived from the combination of the Montecarlo simulations, cosmic variance and
Poissonian errors. Black dotted line: local MF \citep{Cole2001}. Dashed vertical
lines represent the completeness in mass for each redshift bin. GSMFs from other authors 
are also plotted: blue triangles from \citet{PG2008}, green
astrerisks from \citet{Fontana2006}, red diamonds from \citet{Marchesini2009},
orange squares form \citet{Ilbert2010}.}
\end{figure*}

Fig. 14 presents the total and the quiescent GSMF  (QSMF hereafter) of our IRAC selected sample
in 4 different redshift bins: 1.4-1.6,
1.6-2.0, 2.0-2.5 and 2.5-3.0. At $z > 3.0$ the uncertainties become dominant so
we decided to limit our analysis to  $z \leq$ 3.0.
The circles represent the GSMF for all the sources, while the lines show
the best Schechter function fitting our data. 
The Poissonian errors have been computed assuming Poissonian statistics
(\citealt{Zucca1997}):

\begin{equation}
\sigma=[\sum_{i}(\frac{1}{V^{2}_{max}(M_{i})})]^{1/2} 
\end{equation}

However, the Poisson errors in our GSMF are an under-estimate of the real
uncertainties, since they do not take into account how the $z_{p}$ uncertainties
propagate on the GSMF. This uncertainty is especially important for the
optically undetected sources, therefore  we have to be very careful in the $z$-bins
where these sources dominate (higher redshift).
To assess for this uncertainty we have performed Montecarlo simulations. We have
created 20 mock catalogues by randomly picking a redshift within the redshift
probability distribution function (PDFz) of each object (given by \textit{Le
Phare} code), we have recalculated the mass and other important physical parameters
 and we have computed the GSMF for each of the 20 mock catalogues. We have
then calculated the $1\sigma$ dispersion of the GSMF values at each mass and
redshift bin. The cosmic variance is another important source of uncertainty. Following the formalism of 
\citet{Somerville2004} we have derived $ \sigma_{cv} \sim 0.18,0.14,0.14$ and $0.16$ respectively in the four redshift bins.
Finally, we added in quadrature these errors to the Poissonian
errors (yellow area in Fig. 14).
The red area represents the upper and lower limits of the QSMF
calculated in the same way as for the total GSMF.

\subsection{Comparison of the total GSMF with literature}

In Fig. 14 we show for comparison the GSMF derivations from the literature (blue
triangles from \citealt{PG2008}, green asterisks from \citealt{Fontana2006}, red
diamonds from \citealt{Marchesini2009}, orange squares from
\citealt{Ilbert2010}). All GSMF's were converted to Chabrier IMF and shifted by $-0.14$
dex in mass (to account for the difference in the derived mass when using
\citealt{Maraston2005} or \citealt{BC2003} models, see \citealt{Pozzetti2010}). We are in reasonably good  agreement with
many of the other results.
However, a significant difference is observed in all redshift bins at high masses 
 with respect to the results of Pérez-González et al. (2008), as they found a number density significantly higher than ours. 
With respect to \citet{Marchesini2009}, we find less massive galaxies at $z <$ 2 and more intermediate mass 
($logM \sim 10.5 $) galaxies at $2 < z < 2.5$. 
The agreement with \citet{Ilbert2010}, whose sample is also IRAC selected  in the COSMOS field, is very good, with small
differences probably arising from different redshift binning, as it may happen also with
\citet{Fontana2006}.
However, we have a slight overdensity of intermediate mass galaxies ($M\sim 10^{10.7}$M$_{\odot}$) at $2.0< z< 2.5$
when compared with all other results from literature, which may be too big to be
explained only by differences in redshift binning. 

In comparison with other authors, we are using data from the COSMOS survey,
which is the largest studied area ($ \sim1.7$ deg$^{2}$) with full multi-wavelength coverage.


\subsection{The evolution of the quiescent galaxies}

\begin{figure}
\centering
\includegraphics[angle=0,
width=0.5\textwidth]{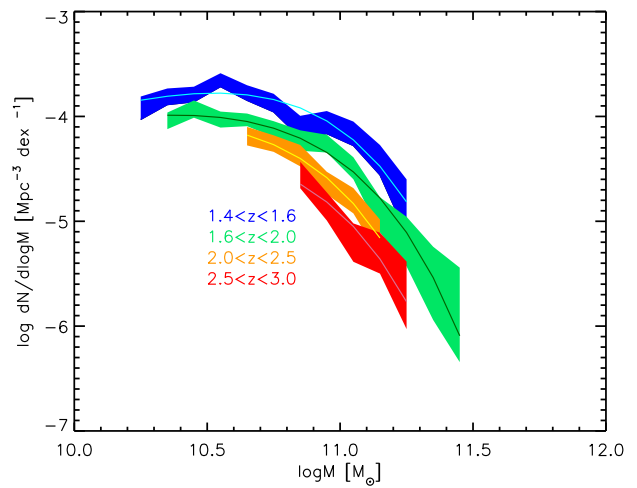}
\caption{Evolution of the QSMF with redshift for different redshift bins. Blue:
$1.4-1.6$; green: $1.6-2.0$ ; orange: $2.0-2.5$; red: $2.5-3.0$. Shaded regions
represent the upper and lower limits of the GSMF for each redshift bin (see Section 7),
 while the thick lines are the Schechter fitting to the data.}
\end{figure}

In Fig. 15  the QSMF obtained in the four studied redshift bins is reported. The shaded areas represent the 
upper and lower limits, derived as explained in Section 7, while the thick lines are the Schechter fitting to our data.

We observe a significant evolution of the QSMF from $2.5 < z < 3.0$ to  $  1.4< z < 1.6$, increasing $\sim 1$ dex 
for galaxies with $log M \sim 11.0$ in the redshift interval studied. The evolution is  $\sim 0.3$ dex 
in each redshift bin, meaning that the number of quiescent galaxies continously increases with cosmic time.

In Fig. 16 we compare the GSMF derived for quiescent galaxies, with those of the other populations.
The red, green and blue areas represent the quiescent, intermediate and actively star-forming populations, respectively.
We also show for comparison the results obtained by \citet{Ilbert2010}, with
purple, dark green and light blue circles representing their quiescent,
intermediate and high activity galaxies respectively. The agreement is
 good for the quiescent population in both redshift bins in common ($1.4-1.6$, $1.6-2.0$), while the
differences for the intermediate and star-forming population can be explained
by the different classification methods used. While for the quiescent population we used  a similar cut in SSFR,
 for the cut between the intermediate and high activity populations they used $log(SSFR [$Gyr$^-1])\sim -0.3$, while we are using a value of $-0.5$.

The evolution of the star-forming galaxies is quite complex (see Fig. 16). They remain almost constant from
$2.5 <z < 3.0$ to $2.0< z< 2.5$, being the dominant population  at all masses in those redshift bins.
However, their number density decreases when moving to the lowest redshift bins and this decrease depends on the galaxy mass: while for 
high masses  we observe a significant decline ($\sim$ 0.8 dex for $logM \sim 11.0$), for low masses  ($logM \sim 10.5$) the GSMF is almost constant.
 A decrease of the star-forming population from $1.6< z< 2.0 $ to  $1.4 <z < 1.6$ is in agreement with
the results of \citet{Ilbert2010} and, as reported in studies at lower redshift (see for example \citealt{Ilbert2010}, \citealt{Pozzetti2009}),
the number of star-forming galaxies continues to decrease to almost local redshift ($z \sim 0.1$).
The intermediate population constantly increases its number at all masses, first at high masses ($\sim 0.2$ dex from $2.5 <z < 3.0$
 to $1.6< z< 2.0 $ for $logM \sim 11.0$), then at low masses ($\sim 0.2$ dex from $1.6< z< 2.0 $ to $1.4 <z < 1.6$ for $logM \sim 10.5$).
This means that they become more abundant than the star-forming population at $logM \gtrsim 10.6$ at $1.4 <z <1.6$.
Finally, as mentioned before, for the quiescent population we find an increase of $\sim 1$dex in the number density  
 from $2.5 <z < 3.0$  to  $1.4 <z< 1.6$ for $logM \sim 11.0$.
At the lowest redshift bin ( $1.4 <z <1.6$) the quiescent population becomes more important than the active one for masses $log M \gtrsim 10.75$.

Considering all the populations, we find that $z \sim1.5$ is a clear epoch of
transition of the GSMF: while the GSMF at $z \gtrsim 1.5$ is dominated by the
actively star-forming galaxies at all stellar masses, at $z \lesssim1.5$ the
contribution to the total GSMF of the quiescent galaxies is significant and
both the intermediate and the quiescent population become more important than the star-forming
 one for $M \geqslant 10^{10.75}$M$_{\odot}$.

\begin{figure*}
\centering
\includegraphics[angle=0,
width=1.0\textwidth]{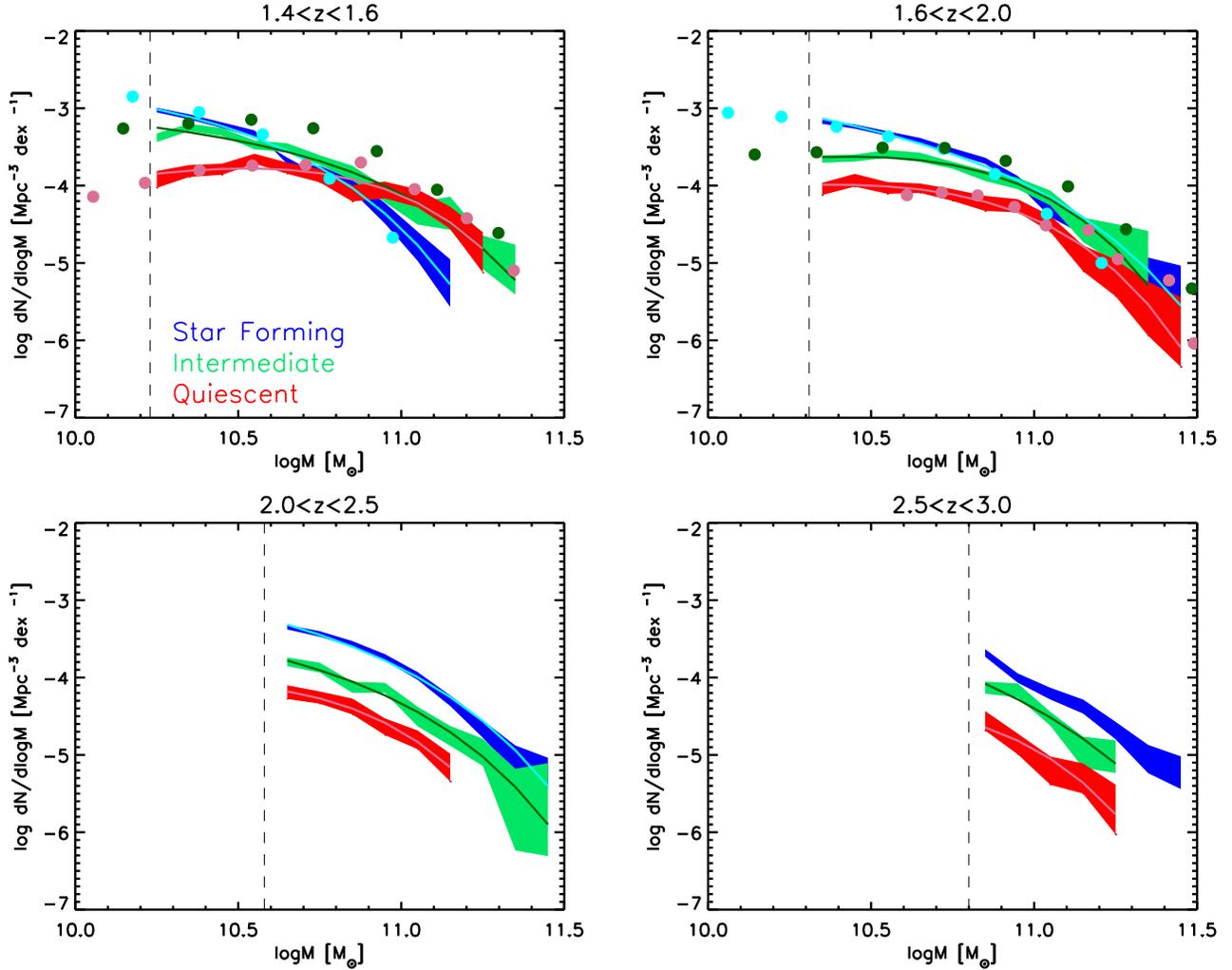}
\caption{Evolution of the GSMF for different activity types galaxies. The
colored areas represent the upper and lower limits (see Section 7) while the thick lines are the Schechter fitting to the data.
Red: quiescent. Green: intermediate. Blue: actively star-forming. The circles
represent the GSMF for quiescent (purple), intermediate (dark green) and high
 activity galaxies (light blue), as obtained by \citet{Ilbert2010}. Also shown are
the limits in mass (vertical dashed lines).}
\end{figure*}

In Table 5 we report the best-fit Schechter parameters for the quiescent sample. In the last redshift bin ($2.5 <z < 3.0$ ) the
Schechter fit failed to converge when leaving all the parameters free. We fixed the faint end slope $\alpha$ to its
value at $2.0 <z < 2.5$ to be able to obtain the other Schechter parameters.
While, given the large uncertainties, we cannot derive firm conclusions on the faint end slope $\alpha$
nor on  $M^{*}$, we detect, as already mentioned, a significant increase of $\phi$ from $2.5 <z < 3.0$  to  $1.4 <z < 1.6$.

\begin{table*}
\begin{tabular}{|l|c|c|c|c|}
 \hline
Type        & z-bin & $\alpha$ & $log(M^{*}) [$M$_{\odot}]$     & $\phi [10^{-4}$Mpc$^{-3}$ dex$^{-1}]$\\
\hline\hline

%
Quiescent & 1.4-1.6 & -0.08$\pm$ 0.98  & 10.57 $\pm$0.25 & 1.9$\pm$ 0.5\\
         &  1.6-2.0 &  -0.41$\pm$ 0.92 & 10.61 $\pm$0.19 & 1.0$\pm$ 0.4 \\
         &  2.0-2.5 &  -0.34$\pm$ 4.83 & 10.50 $\pm$0.84 & 0.9$\pm$ 0.8\\
          & 2.5-3.0 &  -0.34 (Fixed)   & 10.52 $\pm$0.18 & 0.5$\pm$ 0.6\\
\hline
%

\end{tabular}
\caption{Schechter parameters of the QSMF at $1.4 \leq z \leq 3.0$.}
\end{table*}

We caution the reader that the uncertainties of this derived GSMF at such high
redshift may change our results significantly due to the difficulties in
estimating the $z_{p}$ as well as the activity classification.

\section{Discussion}

To better understand the behaviour of the different populations we have plotted
the fraction of quiescent/star-forming galaxies over the total versus  mass
for different redshift bins (Fig. 17). The fraction of the actively star-forming 
galaxies decreases from $\sim 60\%$ to $\sim 20\%$ from
$ 2.5< z<3.0$ to $1.4< z < 1.6$ for  $M \gtrsim 10^{11.0}$M$_{\odot}$, while that of quiescent
galaxies increases from $\sim 10\%$ to $\sim 50\%$ in the same redshift and mass intervals. The
evolution is also mass-dependent, with the most massive galaxies becoming
quiescent first, while at $M\lesssim 10^{10.5}$M$_{\odot}$ the star-forming population
still dominates the sample at all redshifts.

This trend suggests that actively star-forming galaxies, which dominate the
sample at high redshifts (mainly at  $z \gtrsim 2.0$), may quench their star
formation and become less abundant at later time.
This quenching of the star formation seems to be mass dependent, with the most massive galaxies quenching their star-formation first,
while the less massive galaxies continue to form stars until later times.
If the evolution of the star-forming galaxies should be that of becoming  intermediate galaxies to finally end their lives as
quiescent galaxies, the increase of this latter population at low redshifts
could be explained, with the quiescent population being the dominant population
at low redshift, as all galaxies may evolve to that type once their SF has stopped. 
Therefore, we could be unveling the formation epoch of the very active galaxies
and their following evolution from the blue cloud to the red sequence.

\begin{figure}
\centering
\includegraphics[angle=0,
width=0.5\textwidth]{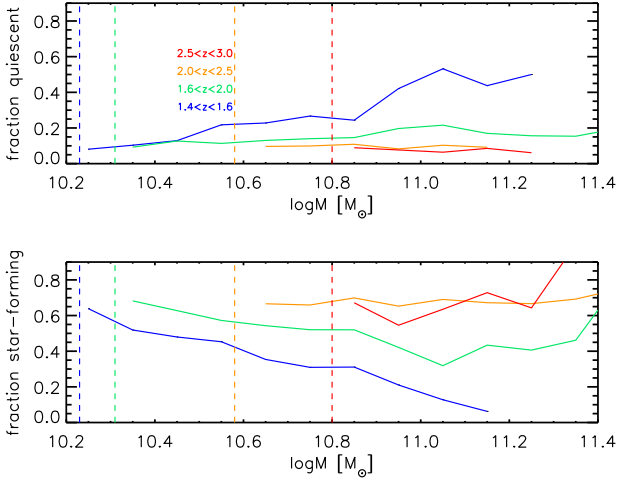}
\caption{Upper panel: Fraction of quiescent galaxies over total for different
mass values. Colors represent the different redshift bins as in Fig. 15. Lower panel: Fraction of 
the star-forming population over total versus mass. Colored vertical dashed lines are the mass
limits for each redshift bin.}
\end{figure}

\begin{figure}
\centering
\includegraphics[angle=0,
width=0.5\textwidth]{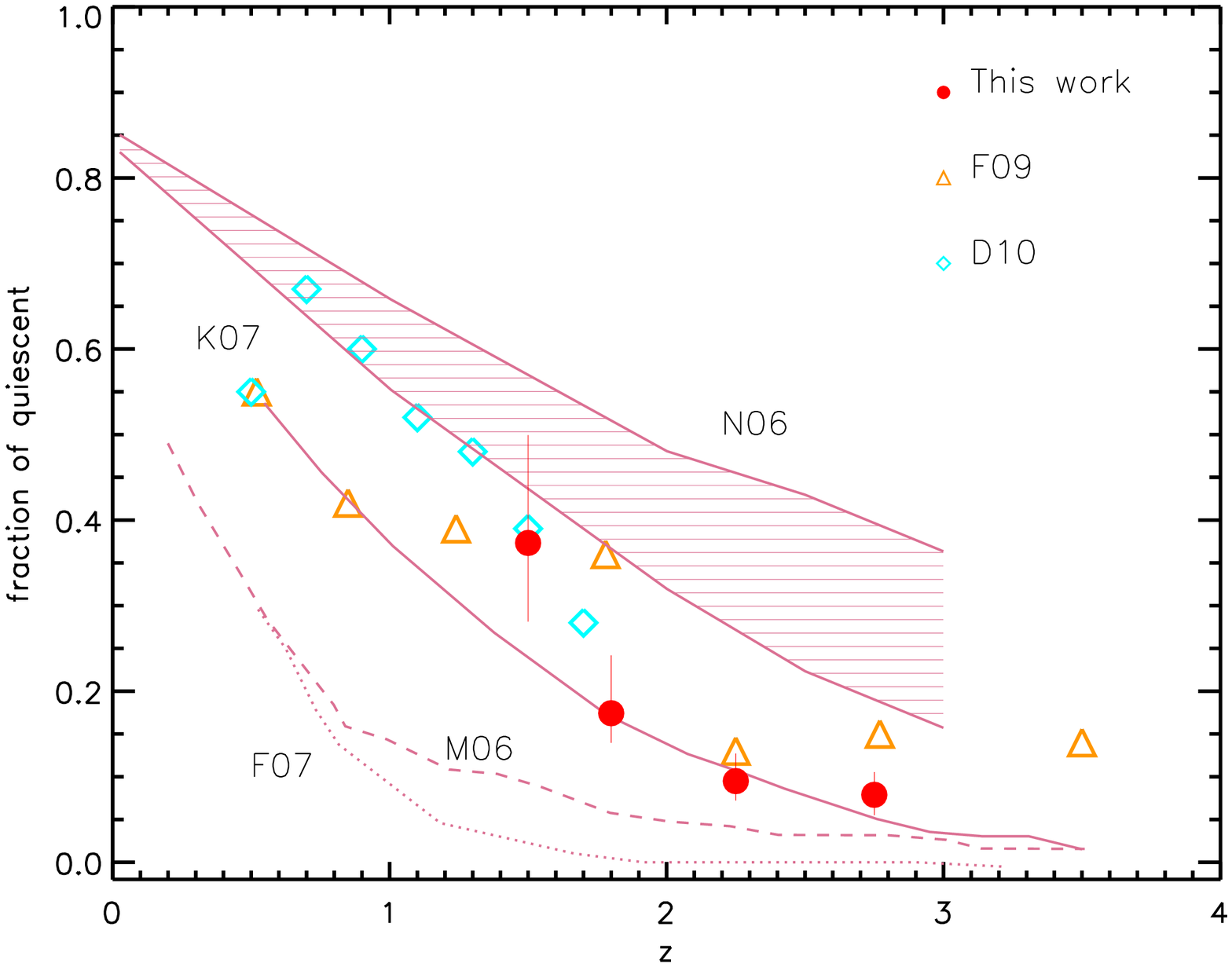}
\caption{Fraction of quiescent galaxies ($log(SSFR [$Gyr$^1])<-2$)  with $ M \geq
10^{10.85}$M$_{\odot}$ as a function of redshift. Red circles are the results
obtained in this work (errors are the combination of the Poisson errors and the
Montecarlo simulations), orange triangles are the values from \citet{Fontana2009}
and light blue diamonds from \citet{Damen2010}. Lines refer to the predictions of
theoretical models as decribed in the legend: \citet{Menci2006} (M06),
\citet{Kitzbichler2007} (K07), \citet{Nagamine2006} (N06) and \citet{Fontanot2007}
(F07).}
\end{figure}

Fig. 18 shows the evolution of the fraction of quiescent galaxies with redshift
for sources with $M\geq10^{10.85}$M$_{\odot}$ (red circles). The errors are a
combination of the Montecarlo simulations and Poissonian errors. For comparison, we also plotted
the values obtained by Fontana et al. (2009, orange triangles) and from
Damen et al. (2010, light blue diamonds), as well as the predictions of theoretical models
(\citealt{Menci2006}, M06 hereafter; \citealt{Kitzbichler2007}, K07; \citealt{Nagamine2006}, N06; 
\citealt{Fontanot2007}, F07). We observe a good agreement between the fraction of
quiescent galaxies derived in this work and those from \citet{Fontana2009}. 
 The values of \citet{Damen2010} are also consistent with our
results, although \citet{Damen2010} sources are selected with  $M > 10^{11.0}$M$_{\odot}$ and are
defined as quiescent when $SSFR < 1/(3 t_{H})$. 
The fraction of massive quiescent galaxies increases from $\sim 8\%$ to $\sim 35\%$ from $2.5< z < 3.0$ to $ 1.4< z<1.6$,
with the main evolution occuring between $1.6< z < 2.0$ and  $1.4< z < 1.6$ (where the fraction increases from  $\sim 18\%$ to $\sim 35\%$).
This confirms the expected cosmological increase in number density of massive quiescent galaxies with cosmic time. All the theoretical models
agree in predicting a gradual increase with time in the fraction of galaxies
with low SFR, although large discrepancies between data and theoretical models are observed.
It is beyond the scope of the present paper to give a detailed explanation of the differences between models, 
but we will discuss some generalities. 
Some models underpredict the fraction of quiescent galaxies at all redshifts, such as the purely semi-analytical models (M06, F07), 
and in particular predict virtually no such objects at $z > 2$, in contrast to what is observed. This lack of massive galaxies at
high redshift for the model of F07 has already been noticed by \citet{Fontanot2007}, who admitted that the 
downsizing trend of the galaxies is not fully reproduced yet in these models, and suggested that some kind of feedback mechanism
could help to reproduce the data. A slightly higher fraction of quiescent massive galaxies is predicted by M06, as they include 
AGN feedback mechanism, due to the growth of supermassive black holes and the AGN triggered by interactions in the host galaxies.
This AGN feedback enhances the fraction of galaxies populating the red branch of the color distribution and is particularly effective at high z.
However, it is not effective enough to reproduce the observed data at $z < 2.0$. 
The purely hydro-dynamical model by \citet{Nagamine2006}, represented with a shaded area for three different timescales $\tau$ of the star-formation rate 
(ranging from 2 x $10^{7}$ yrs to 2 x $10^{8}$ yrs), appears to overpredict the fraction of high mass quiescent galaxies. 
The disagreement in the stellar mass density at $z > 1$ between these models and the observations has already been commented by \citet{Nagamine2006}.
 The semi-analytical rendition of the Millennium N-body dark matter simulation by \citet{Kitzbichler2007} is the only one which  agrees with 
our observed data, being able to reproduce both the normalization and  the shape of the increase in the fraction of quiescent massive galaxies
 with decreasing redshift. This model introduces radio mode feedback from the central galaxies of groups and clusters, which seems to be 
fundamental for the predictions to be consistent with the observations.

Finally, the non negligible fraction of quiescent galaxies  at $z > 1.8$ implies that these galaxies
assembled most of their stellar mass either during an active star-burst
phase or through important merging processes at higher redshifts. These star
formation episodes must be quenched either by efficient feedback mechanism
and/or by the stochastic nature of the hierarchical merging process.

\begin{figure}
\centering
\includegraphics[angle=0,
width=0.5\textwidth]{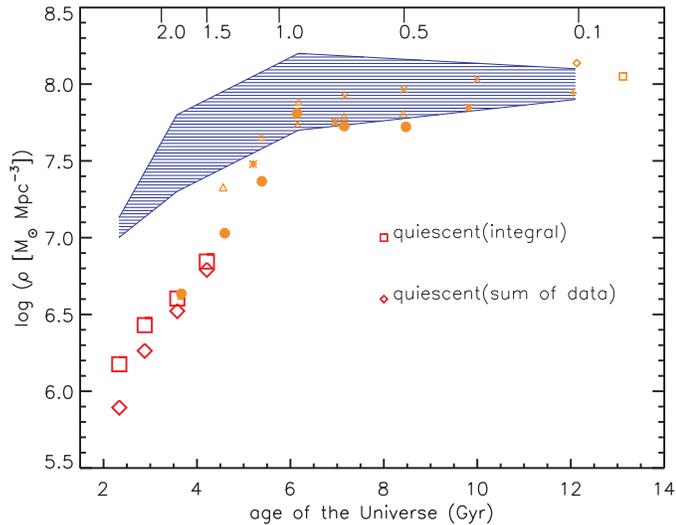}
\caption{Evolution of the stellar mass density of quiescent galaxies. The result from this work are the
red symbols. The values represented by the squares have been
calculated as the integral of the Schechter function fitting the data, while the diamonds
are a mere sum of the data for all masses. Orange  symbols are
results from literature as explained in the article . We also show for comparison the evolution of the stellar mass density 
of the star-forming plus the intermediate galaxies (blue shaded area). }
\end{figure}

\begin{figure}
\centering
\includegraphics[angle=0,
width=0.5\textwidth]{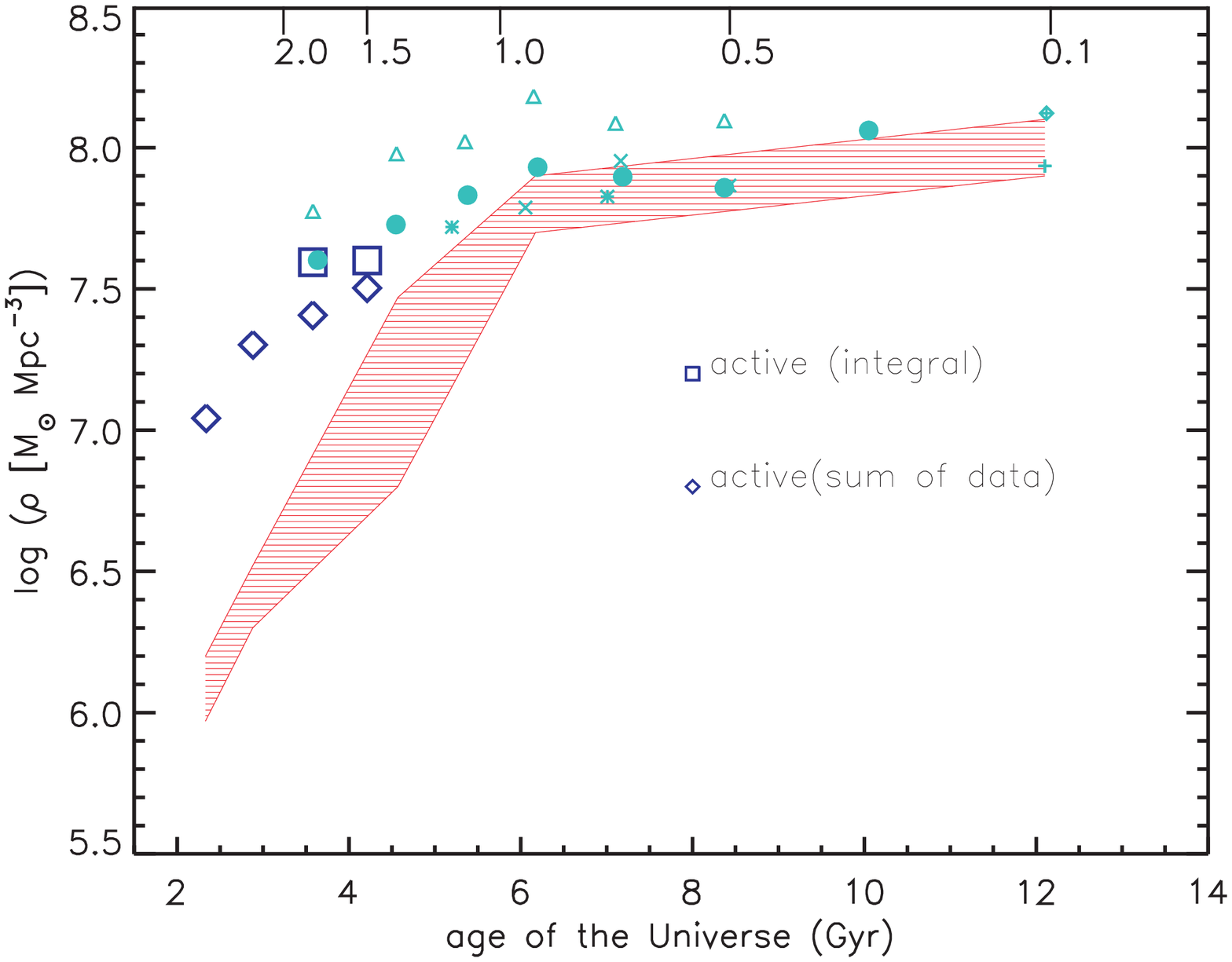}
\caption{Evolution of the stellar mass density of star-forming plus quiescent galaxies. The result from this work are the
dark blue symbols. The values represented by the squares have been
calculated as the integral of the Schechter function fitting the data, while the diamonds
are a mere sum of the data for all masses. Light blue  symbols are
results from literature as explained in the article. We also show for comparison the evolution of the stellar mass density 
of the quiescent galaxies (red shaded area). }
\end{figure}

In Fig. 19 and 20  we have calculated the evolution of the stellar mass density with
cosmic time for the quiescent sample (red symbols) and the intermediate plus 
the star-forming samples (dark blue). The values have been obtained in two different ways. The data
plotted as squares represent the integral of the Schechter function fitting our
data, including the extrapolations to the fainter masses. The diamonds are the sum of
the data at all masses.  As expected, the latter values are lower (since data are
incomplete at low masses) but the two estimates are consistent within $ \sim 0.18$ dex.
We do not show the value of the stellar mass density for the intermediate and star-forming 
galaxies in the redshift bins $2.0-2.5$ and $2.5-3.0$ derived through the integration of the Schechter parameters, neither 
 the same value at $2.5< z< 3.0$ for the quiescent galaxies, due to the high uncertainties in the derived parameters given the
 incompletness at low masses.
In the plots we also show a colored area representing the upper and lower limits of the quiescent (red) and star-forming (blue)
populations, as well as various results from literature. Orange symbols represent the stellar mass density of the quiescent galaxies 
as derived by other authors, while light blue symbols stand for the star-forming ones (squares from \citealt{Kochanek2001}, diamonds from
 \citet{Driver2006}, plus symbols from \citet{Bell2003}, triangles from \citealt{Arnouts2007}, asterisks from \citealt{Franceschini2006},
 crosses from \citet{Borch2006}, dots from \citealt{Ilbert2010}).
We are in very good agreement with Ilbert et al. (2010) in the common redshift bins. 
We observe that the increase in stellar mass density in the redshift range
of interest ($t \sim 2-4.5$ Gyr) is very rapid for both populations ($\sim 0.5$ dex for the star-forming plus intermediate populations
and $\sim 0.9$ dex for the quiescent one), then slowing down at lower redshifts. However, this slow down happens earlier
in time for the star-forming population ( $z \sim 1.2$, $t \sim 5.0$ Gyr) than
for the quiescent galaxies, which continue to rapidly assembly  mass  until later
times ($z \sim 1.0$,  $t \sim 6.0$ Gyr ), reaching the star-forming galaxies mass density at $z < 1.0 $.

The existence of a non negligible population ($log\rho[$M$_{\odot}$ Mpc$^{-3}]$ $\sim 6.0$ at $z\sim 2.7$)
of quiescent galaxies with high masses which have already undergone major star
formation even at the highest redshifts ($z > 2.5$) is fundamental for our
understanding of the galaxy formation processes and crucial for testing
theoretical scenarios.

\section{Summary and Conclusions}

We derived the GSMF and stellar mass density in a  1.73 deg$^{2}$ area in the
COSMOS field from $z=1.4$ to $z=3.0$ for the whole sample and for the sample divided
into populations, paying special attention to the quiescent one.
 The GSMF estimate is based on $\sim$ 18000 galaxies with $mag_{3.6 \mu m}  < $22.0 and $z_{p} \geq 1.4$. 
We summarize our results below:
\begin{enumerate}

 \item  We find that an IRAC selected sample is fundamental for studying
quiescent galaxies at high redshift. About 88$\%$ and 91$\%$ of the \textit{$K_{s}$} and IRAC
sub-samples  are found  at  $z \gtrsim 1.4$ respectively, with this fraction dropping to
21$\%$ for an \textit{$i^{+}$}-band selected sample.
Considering the galaxy classification, 62$\%$ of the optically selected sub-sample is fitted by
a blue template, while the same fraction is much lower ($\sim 7\%$) for the IRAC
sub-sample.

\item We study the main properties of the high redshift quiescent sample ($z \geq  1.4$), finding
that they are old and massive galaxies, with $\langle M \rangle \sim 10^{10.65}$M$_{\odot}$ and ages ranging
from $\sim 1$ to $ \sim 4$ Gyr. They have small \textit{e-folding} time scales $\tau \sim$ $0.1-0.3$ Gyr and very low dust
extinction (E(B-V) $\sim 0-0.15$), meaning passively evolving populations.

\item We observe a significant evolution of the QSMF from $2.5 < z < 3.0$ to $1.4 < z < 1.6$,
 amounting to $\sim 1$ dex for galaxies with $log M \sim 11.0$. The evolution is 
$\sim 0.3$ dex in each redshift bin, meaning that the number of quiescent galaxies continously increases with cosmic time.

\item We find that $z\sim1.5$ is a clear epoch of transition of the GSMF: while
the GSMF at $z \gtrsim 1.5$ is dominated by the star-forming galaxies at all
stellar masses, at $z \lesssim1.5$ the contribution to the total GSMF of the
quiescent galaxies is significant and the quiescent galaxies become more important
 than the star-forming population for $M \geqslant 10^{10.75}$M$_{\odot}$.

\item The fraction of star-forming galaxies decreases from $60\%$ to
$20\%$ from $2.5 < z < 3.0$ to $1.4< z< 1.6$ for  $M \sim 10^{11.0}$M$_{\odot}$, while
the quiescent population increases from $10\%$ to $50\%$ at the same $z$ and mass
intervals.

\item All of the theoretical models agree in predicting a gradual increase with cosmic time in the fraction of galaxies
with a low SFR. However, only the K07  Millennium-based model is able to properly reproduce the shape of the data.

\item We find a significant number of quiescent galaxies already in place  at $z > 2.5$ ( $ \rho
\sim$ 6.0 M$_{\odot}$Mpc$^{-3}$), meaning that these galaxies assembled most of
their stellar mass in previous epochs in an active star-burst phase or through
important merging processes at higher redshifts. 

\end{enumerate}

\label{lastpage}

\bibliography{bibliografia}

\end{document}